\documentclass[reprint,amsmath,amssymb,longbibliography,aps, prb]{revtex4-2}
\pdfoutput=1
\usepackage{hyperref}
\usepackage{url}
\usepackage{color}
\usepackage{acro}
\usepackage{chemmacros}
\usepackage[normalem]{ulem}

\newcommand{\beq}{\begin{eqnarray}}
\newcommand{\eeq}{\end{eqnarray}}

\usepackage{graphicx}

\begin{document}

\title{ Simulating Superconducting Properties of Overdoped Cuprates: the Role of Inhomogeneity}

\author{Mainak Pal$^{1*}$\thanks{email for correspondence: mainak.pal@ufl.edu},  Andreas Kreisel$^{2,3}$, W.A. Atkinson$^4$, and P.J. Hirschfeld$^1$}

\affiliation{$^1$Department of Physics, University of Florida, Gainesville, Florida, USA}
\affiliation{$^2$Institut f\" ur Theoretische Physik, Universit\"at Leipzig, D-04103 Leipzig, Germany}
\affiliation{$^3$Niels Bohr Institute, University of Copenhagen, 2100 Copenhagen, Denmark}
\affiliation{$^4$Department of Physics and Astronomy, Trent University, Peterborough, Ontario, Canada K9L0G2}

\begin{abstract}
Theoretical studies  of disordered $d$-wave superconductors have focused, with a few exceptions, on optimally doped models with strong scatterers.  Addressing recent controversies about the nature of the overdoped cuprates, however, requires studies of the weaker scattering associated with dopant atoms.  
Here we study simple models of such systems in the self-consistent Bogoliubov-de Gennes (BdG) framework, and compare to disorder-averaged results using the self-consistent-T-matrix-approximation (SCTMA).
Despite surprisingly linear in energy behavior of the low-energy density of states even for quite disordered systems,  the superfluid density in such cases retains a quadratic low-temperature variation of the penetration depth, unlike other BdG results reported recently.  We trace the discrepancy to smaller effective system size employed in that work.  Overall, the SCTMA performs remarkably well, with the exception of highly disordered systems with strongly suppressed superfluid density.  We explore this interesting region where gap inhomogeneity dominates measured superconducting properties, and compare with overdoped cuprates.
\end{abstract}

\maketitle
\section{Introduction} 
{Many phenomena observed in underdoped cuprates are currently highly debated, including the origin of the so-called competing order phases: charge and spin ordering, pseudogap etc.} Overdoped cuprates, which used to be regarded as relatively simpler have recently attracted considerable interest, {due to a few striking observations of unexpected behavior apparently inconsistent with the Landau-BCS paradigm} \cite{Bozovic2016,Mahmood2019,Li2021}.  
{One observation that has received a great deal of attention is the 2D superfluid density in overdoped LSCO. The Brookhaven group prepared a large sequence of high-quality, epitaxially grown overdoped films\cite{ Bozovic2016}, and showed that in a broad range of doping, the superfluid density decreases %with a linear behavior 
{linearly} as temperature is increased. The authors also found that the zero- temperature superfluid density {$\rho_s$ at various dopings is proportional to the critical temperature $T_c$,
% (see Fig. \ref{fig:Bozovic_figure})
in contrast to BCS theory, which states that the zero temperature $\rho_s$ should not depend on $T_c$ at all in a clean system, but instead be simply proportional to the carrier density (which increases as the system is doped).} 
}Although this quasi-proportionality of $\rho_s(0)$ and $T_c$ can occur in the dirty limit\cite{PhysRevB.87.220507},
the authors ruled out an explanation based on disorder, since the expectation is that  the linear temperature variation of the clean superfluid density is replaced by quadratic behavior in a dirty system\cite{PhysRevB.48.4219} which was not observed in their films down to the lowest measurement temperature of about 3K. 

In a series of papers\cite{PhysRevB.96.024501,PhysRevB.98.054506,PhysRevResearch.2.013228,https://doi.org/10.48550/arxiv.2206.01348} based on the so-called ``dirty $d$-wave" theory, David Broun and collaborators, including one of the present authors, have argued that these data can be explained naturally and simply by accounting for the weak scattering nature of the dopant impurities, located away from the CuO$_2$ planes, as well as the correct band structure of LSCO.  
In this approach, the disorder averaged self-energy of a $d$-wave superconductor is calculated using the self-consistent T-matrix approximation (SCTMA).   These authors showed that a very good accounting of the unusual behavior of the overdoped films could be obtained within this Landau-BCS paradigm, without resorting to more exotic explanations.

The SCTMA assumes that disorder is distributed randomly, and replaces the dirty superconductor with an effective translationally invariant medium with dissipation.  However, there is considerable evidence that at least some overdoped cuprate samples are quite inhomogeneous\cite{Tranquada2022,MilanAllan2022} at the nanoscale.  The origin of this inhomogeneity is not completely clear. It may arise  in the sample fabrication and annealing process from chemical barriers, in which case more refined annealing protocols might remove much of the inhomogeneity.  Such an explanation is indeed supported by the claimed homogeneity of the epitaxially grown superconducting films\cite{Bozovic2016} relative to the samples in Refs.\cite{Tranquada2022,MilanAllan2022}.

On the other hand, it is possible that some of the inhomogeneity is emergent, i.e. arises in any short coherence length $d$-wave system with random disorder.  Recently, Li et al.\cite{Li2021} studied the superfluid density vs. doping for a model of a $d$-wave superconductor using BdG simulations, showed that very inhomogeneous gap distributions occurred for large concentrations of weak-to-intermediate impurity potentials, and proposed that an emergent network of $d$-wave superconducting islands connected by weak Josephson links was present in these samples.  Such a description of the superconducting state would perforce lead to large phase fluctuations, which are accounted for neither in the SCTMA nor the BdG calculations, but which might give rise to a low-temperature linear superfluid density behavior\cite{Benfatto2001}. Interestingly, in the supplementary material of the Li et al.  paper\cite{Li2021}, the superfluid density was calculated {\it without} phase fluctuations, and nevertheless 
shown to be remarkably linear in $T$ even for substantial concentrations of moderate strength impurities, which is surprising, given the expectations from the dirty $d$-wave theory.

In this paper, we compare the results of the disorder averaged theory of a $d$-wave superconductor with the  self-consistent BdG method.  The latter is subject to finite-size effects that need to be carefully controlled, but includes both a) scattering processes not included in the SCTMA, and b) the effects of inhomogeneity. Our goals are to test the validity of the SCTMA, which has proven so successful in explaining the properties of the overdoped cuprates\cite{PhysRevB.96.024501,PhysRevB.98.054506,PhysRevResearch.2.013228,https://doi.org/10.48550/arxiv.2206.01348}, and to check the predictions of Li et al.\cite{Li2021} to see if alternative disorder/inhomogeneity based explanations are possible.   A recent paper by Brei{\o} et al.\cite{Breio2022} investigated some of the latter issues, and concluded on the basis of similar BdG calculations that an effective model of weakly-coupled $d$-wave grains did {\it not} emerge naturally from the highly disordered regime.  Here we study 
density of states, superconducting order parameter,  and superfluid density {of a disordered $d$-wave superconductor}, and show that the SCTMA is remarkably robust in explaining the regime of intermediate strength disorder appropriate for dopant atoms, until regions of vanishing superconductivity begin to appear in the sample, either for very high disorder levels or very close to the critical temperature. 

The quasiparticle density of states in this regime is found to be remarkably linear at low energies until the gap is nearly filled.  Despite this, the penetration depth retains its quadratic behavior for disorder levels that strongly suppress the zero temperature superfluid density.  This result, which contradicts that of Li et al.\cite{Li2021}, is obtained properly only for sufficiently large system size and configuration averaging. We note their  result for the $T$-dependence of $\rho_s$ was { not} one of the main points of their paper, nor does this remark affect their main conclusions; however it is interesting and reassuring that the expected results from the SCTMA are indeed accurate.  We note further that the results reported in the present work do {\it not} contradict the result of Ref. \cite{PhysRevB.96.024501}, which stated clearly that the quasi-linear behavior in $\rho_s(T)$ was due in large part to the particular electronic structure of LSCO, which we do not consider here.

\section{Model}
We consider a Hamiltonian for disordered cuprates which assumes weak pointlike potential impurities of a fixed strength randomly at different lattice sites. 
Following Ref. \cite{Li2021}, in order to avoid issues related to the pseudogap, the homogeneous system is taken at optimal doping of $p=$15\%, corresponding to $n=0.85$ electron density.   Disorder introduces an additional $n_\mathrm{imp}$ impurities, each of which dopes the system with approximately one electron, such that $p=0.15+0.5n_\mathrm{imp}$. 
The Hamiltonian is given by
\begin{equation}
    \begin{split}
        \hat  H = \sum_{ij\sigma} \Big\{t_{ij} + \Big(w_{i}- \mu\Big)\delta_{ij}\Big\} c_{i\sigma}^{\dagger} c_{j\sigma} + \hat H_{\mathrm{pair}}
    \end{split}
    \label{eq:BdG}
\end{equation}
where $t_{ij}$ are the tight-binding hopping integrals,  $w_{i}$   are impurity potentials at site $i$ and $\mu$ is the chemical potential. The pairing is treated in a mean-field approximation, 
\begin{equation}
\hat H_{\mathrm{pair}} = \sum_{\langle i,j\rangle}\Big \{\Delta_{ij}c_{i\uparrow}^{\dagger}c_{j\downarrow}^{\dagger} + \mathrm{h.\ c.}\Big\}    
\end{equation}
with $\Delta_{ij}=V\langle c_{i\uparrow}c_{j\downarrow}\rangle$ as the order parameter. A local d-wave order parameter $\Delta_{i}$ can be defined as $\Delta_{i} = \dfrac{1}{4}\big( \Delta_{i,i+\hat{x}} + \Delta_{i,i-\hat{x}} - \Delta_{i,i+\hat{y}} - \Delta_{i,i-\hat{y}}\big)$.
We take $t_{ij} = -1$ for nearest neighbor $i,j$; and $t_{ij} = 0.35$ for next nearest neighbor $i,j$; and zero otherwise, corresponding to a momentum space band for the clean system of $\xi_{\bf k}=t(\cos k_x+\cos k_y)+ t' \cos k_x\cos k_y$.  This ratio of $t/t^{\prime}$ is quite typical for cuprate bandstructures. The pairing interaction is taken as $V = 0.8t$. {In  the clean limit this gives a superconducting transition temperature $kT_c = 0.085t$ when the electron density is $n=0.85$ per unit cell.  This electron density corresponds to a hole density $p=1-n = 0.15$ that is commonly associated with optimal doping in cuprates.}  The corresponding $d$-wave gap function is
$\Delta_k=\Delta_0(\cos k_x -\cos k_y)$, {with $\Delta_0=2\Delta_i = 0.0935{t}$}.  Note that since in cuprates $t\sim$ 300 meV from ARPES, $T_c\sim 300$ K, which is artificially high. The parameters are chosen in this way to allow  a direct comparison with Ref.~\cite{Li2021}, and for numerical convenience.  This amounts to expressing  quantities with  dimensions of energy in units of $t$ without explicitly mentioning the unit.
\begin{figure*}[tb]
    \centering
    \includegraphics[width=1\linewidth]{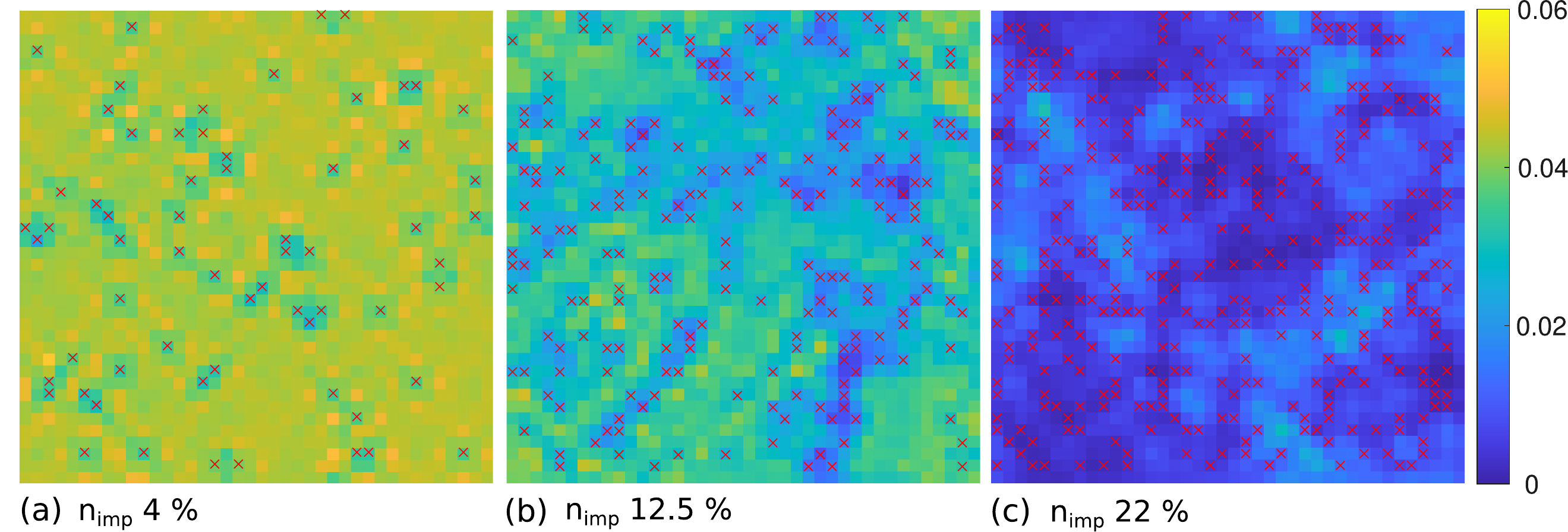}
    \caption{Maps of typical random impurity configurations and corresponding superconducting d-wave order parameters $\Delta_i$ (colorscale) at three  impurity concentrations: (a) $4 \%$ (b) 12.5\% (c) 22\% at $kT/t=0.001$. Impurity locations are indicated by red crosses.  The impurity potential is $1t$.
    }
    \label{fig:gapmaps}
\end{figure*}

The impurity { potential} $w_{i}$ at any site $i$ is chosen randomly { to be} either 0 or 1. A scattering potential of 1 ($t$) amounts to a  moderate strength of the potential scatterers we are modeling, well away from the unitarity limit, but still considerably larger than the realistic ab initio dopant potentials of order 0.2$t$--0.3$t$ calculated in Ref.~\cite{https://doi.org/10.48550/arxiv.2206.01348} for Sr dopants in LSCO.  In this sense our study is qualitative, and makes no attempt to compare to real cuprate systems. We have also not attempted to include O vacancies and other sources of disorder discussed in Ref.~\cite{https://doi.org/10.48550/arxiv.2206.01348}.
The chemical potential $\mu$ is determined self-consistently via iterative diagonalization of the BdG Hamiltonian. Fixing the average electron density in the system leads to a unique chemical potential via the self-consistent diagonalization.
\section{Results}
We work in a $40\times40$ lattice with periodic bondary condition, and {average over} several random impurity configurations at various values of average impurity density $n_{\text{imp}}$.
To increase the resolution of the spectrum and for the purpose of studying the system at sufficiently  low temperature, the supercell method  is employed to ensure a dense set of energy levels (for details see {Appendix \ref{supercell_method}}).
With a band-width of {$\sim 8t$} and only $40\times40$ levels, the level-spacing may not be sufficient at low energies if supercells are not used. 
Averaging over random configurations of impurities was also used to improve the statistics. Averaging over more and more impurity configurations makes the system effectively more and more translationally invariant. Typically, averaging over 40 random configurations of impurities appears to be sufficient to get accurate low-energy behavior.

The BdG equations are iteratively and self-consistently solved for the order parameter and chemical potential.
If the impurity concentration in the system is varied  with fixed chemical potential, doping varies proportional to the impurity concentration, and may be compared\cite{Li2021} roughly to a cuprate scale as  $p=0.15+n_{imp}/2$. Note, however, that the actual effect of doping on the electronic structure is neglected here, and the rate of doping change with impurity concentration depends in the model entirely on the potential of each individual dopant, taken here to be $w=1t$.
Results for $d$-wave order parameter, density of states and superfluid density are summarized below.
\subsection{$d$-wave order parameter}
First, the variation of the superconducting order parameter was studied as a function of the impurity concentration. An impurity concentration of $ 0< n_\mathrm{imp} < 1$ empirically corresponds to a hole doping of roughly $\Delta p \approx 0.5n_\mathrm{imp}$ with respect to the optimal filling of $p=0.15$. We worked within an impurity concentration range of  $n_\mathrm{imp} =0.02$--$0.3$.
Typical maps of impurity configurations along with the $d$-wave order parameter at moderate to high impurity concentrations are shown in Fig. \ref{fig:gapmaps}.  It is clear for the highest concentration that impurities of this strength suppress the superconductivity nearly completely, and that the resistive transition temperature, which depends on the existence of a percolating path of nonzero order, might be zero.{
}{
}Note that it is known that electrostatic potentials of this kind cannot reproduce the detailed statistics of optimally doped cuprates, which require in addition a disorder component of the Andreev type\cite{Nunner2005}.
Nevertheless it is clear that dopant disorder of this form can lead to extensive gap inhomogneity for higher concentrations\cite{Li2021}.

In these simulations, since the pairing interaction is uniform on all nearest-neighbor bonds, a small local gap component of $s$-wave symmetry will be induced by the disorder.  In addition, higher concentrations will lead to time-reversal-symmetry breaking\cite{Li2021,Breio2022} by local currents, despite the fact that the Hamiltonian  contains no terms that explicitly break time reversal symmetry. 
We study neither of these effects here.  
\begin{figure}[tb]
    \centering
    \includegraphics[width=1\linewidth]{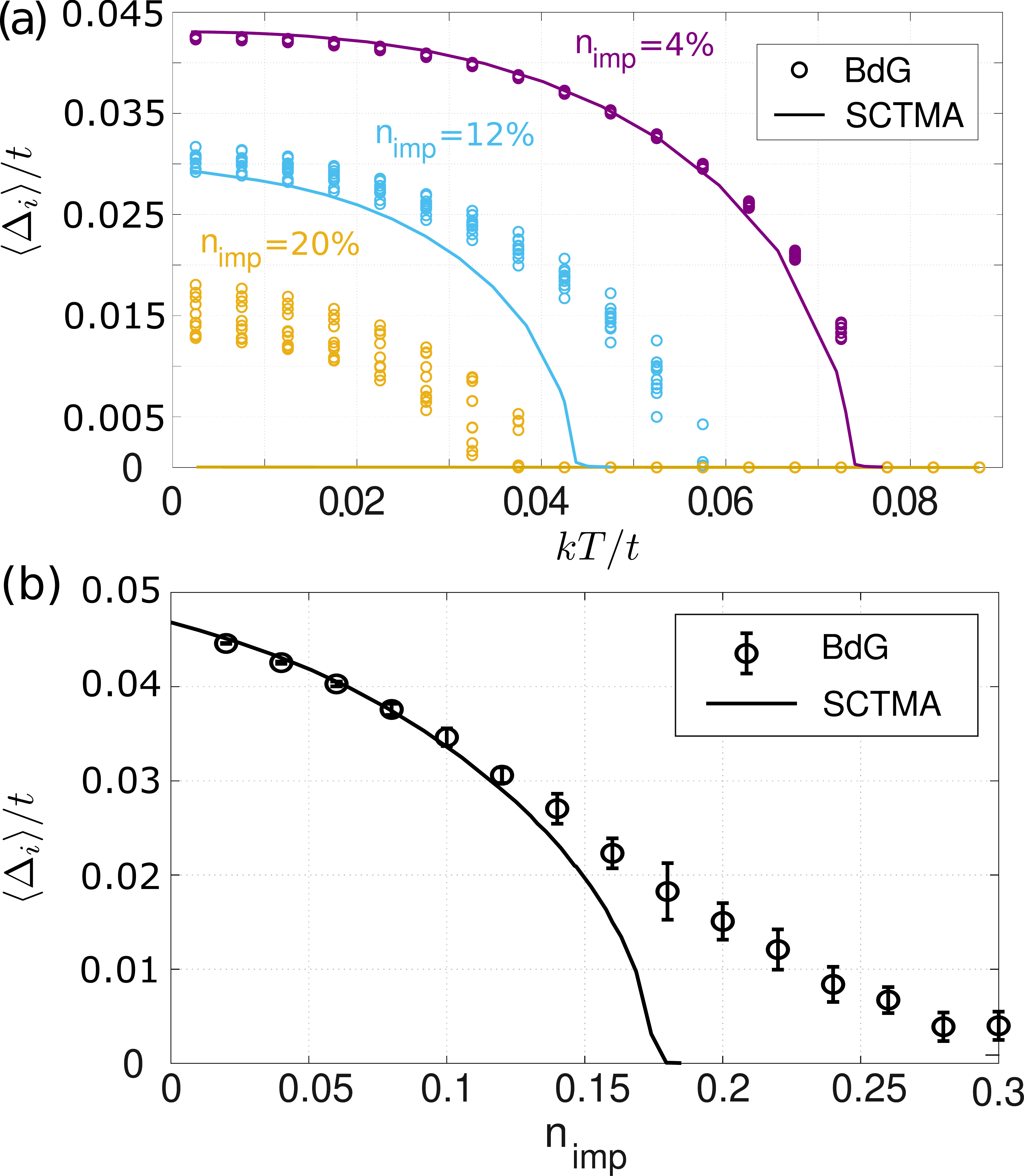}
    \caption{ (a) Plot of projected $d$-wave  bond order parameter $\Delta_i$ averaged over entire system for 10 impurity configurations  (circles), for $n_\mathrm{imp}$=4\%,12\% and 20\% (symbols) and compared to SCTMA (solid lines).  (b) Average order parameter at $T\rightarrow 0$ vs.~impurity concentration $n_{\text{imp}}$. Error bars represent standard deviation over the different configurations.
    }
    \label{fig:avg_op}
\end{figure}

The  order parameter maps shown in Fig.~\ref{fig:gapmaps} hint at another
interesting aspect of the highly disordered state,  namely that the order parameter inhomogeneity increases with $n_\mathrm{imp}$.  Fig. \ref{fig:avg_op}(a) compares the temperature dependence of the average bond order parameter
for a number of disorder configurations; each data point corresponds to the average of $\Delta_i$ over the lattice for a single configuration. Here it is seen that not only does the gap distribution become more inhomogeneous with increasing disorder, it also becomes more sensitive to unusual local disorder configurations\cite{Li2021}, giving rise to a much larger dispersion of average gap values over configurations.  Also shown are the results for the same concentrations and model within the SCTMA (described in Appendix \ref{app:C}).
For the less disordered systems, the SCTMA describes the $T$-dependent order parameter semiquantitatively at low $T$, but predicts gap closings at lower $T_c$s than BdG (where $T_c>0$ only means that $\Delta_i>0$ on as little as 1 site).  For the most disordered case shown, the SCTMA predicts $\Delta=0$ at all temperatures.  The strongly inhomogengeous states found in BdG indeed persist up to much higher doping, as seen in Fig. \ref{fig:avg_op}(b).

\subsection{Density of states}
\begin{figure}[tb]
    \centering
    \includegraphics[width=1\linewidth]{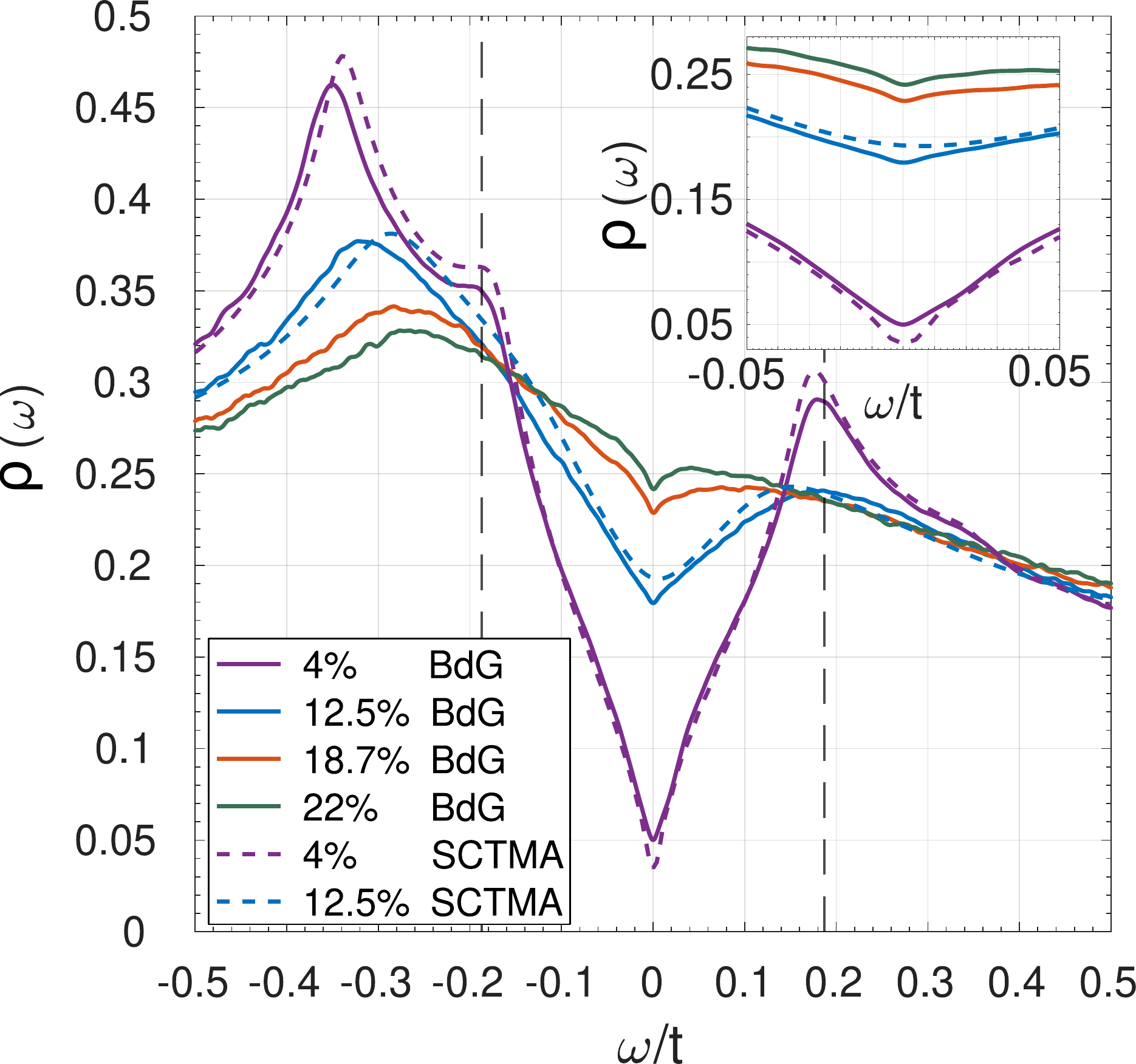}    
    \caption[Density of states $N(\omega)$ of the system at $kT = 0.001$ averaged over all lattice sites of $40$ random impurity configurations (with $15\times15$ supercells)]{Density of states $\rho(\omega)$ of the system at $kT = 0.001$ averaged over all lattice sites of $40$ random impurity configurations (with $15\times15$ supercells) at each impurity concentration 12.5\%, 18.7\% and 22\%.
    Artificial broadening is $\eta/t=0.005$, impurity potential $w_{ii}/t = 1$. In clean homogeneous system, d-wave order parameter $\Delta/t$ is $0.187$ where dashed vertical lines indicate the coherence peak positions in the 4\% case.  Dashed colored lines indicate the corresponding SCTMA results for 4\% and 12.5\%. } 
    \label{fig:LDOS_12pt5_18pt7_22}
\end{figure}

In the  BCS theory of a clean $d$-wave superconductor, the low temperature linear-$T$ term arises from the linear in $\omega$ density of states.  It is therefore interesting to study how the density of states varies with energy in the disordered case with intermediate strength scattering.
{The single-spin local density of states (LDOS) at lattice site $i$ can be calculated 
from the eigenvectors $U_{i\sigma, ns}$ and eigenvalues $E_{ns}$ of the BdG Hamiltonian, Eq.~(\ref{eq:BdG}),
\begin{equation}
    \rho({\bf r}_i, \omega) = -\dfrac{1}{\pi}{\rm Im}\Bigg(\displaystyle{\sum_{n, s}}\dfrac{|U_{i\uparrow ,ns}|^{2}}{\omega-E_{ns}+i\eta^{+}}\Bigg)
\end{equation} 
where $\eta^{+}$ is an artificial broadening, the pair $(i,\uparrow)$ indicates the projection of the eigenstate $(n,s)$ onto lattice site $i$ with spin up. Here, the pair $(n,s)$ labels eigenstates with pseudospin $s = \Uparrow,\, \Downarrow$ such that $E_{n\Uparrow} >0$ and $E_{n\Downarrow} < 0$ (see Appendix~\ref{app:B})}.  While there are several BdG studies with strong impurities\cite{Atkinson2000,Ghosal2000,Atkinson2000B,Yashenkin2001,Atkinson2003}, there are relatively few works to address weaker scatterers in the overdoped region\cite{Garg2008,Li2021}.  From the SCTMA it is expected that the residual $\omega\rightarrow 0$ DOS is exponentially small for low concentrations, but becomes rapidly a larger fraction of the normal state DOS as either the concentration or the strength of impurities is increased\cite{PhysRevB.98.054506}.  The residual value is clearly seen in Fig. \ref{fig:LDOS_12pt5_18pt7_22} to increase rapidly from zero as disorder is added.

The more interesting feature of the LDOS for {four} 
disorder concentrations shown in  Fig.~\ref{fig:LDOS_12pt5_18pt7_22} is the persistent V-shape spectrum at
low energies.  In the strong-impurity case, the low-energy bound states hybridize to form a broad plateau-like impurity band,
also seen {qualitatively}  in the BdG results with the exception of an energy range exponentially small in the Dirac cone anisotropy, $\sim \exp(-E_F/\Delta_0)$\cite{PJH_WAA_Jltp}. 
The positive energy coherence peak in the clean system is at $\omega/t \sim 0.18$ and this does not differ much at the impurity concentration of 12.5\%. The peak shifts towards zero as $n_\mathrm{imp}$ is further increased, however. 
{While the peak feature at negative energies resembles a coherence peak, it is in fact the van Hove singularity in this system, and the coherence peak itself is invisible in the Figure.
}
\subsection{Superfluid Density}
In a tetragonal superconductor, the penetration depth for in-plane screening currents is independent of the direction of current flow within the plane.  It is thus sufficient to calculate the penetration depth for the case in which the magnetic vector potential and current are both aligned with the $x$ axis, i.e.\ $\lambda_{xx}$.
From the Kubo formula, the inverse squared penetration depth is the sum of diamagnetic and paramagnetic contributions, 
\begin{eqnarray}
\lambda^{-2} \equiv \lambda_{xx}^{-2} = 
\dfrac{4\pi e^{2}}{c^2} 
%\langle 
\left ( K^{\rm{dia}}_{xx} + K^{\rm{para}}_{xx}
\right ).
%\rangle_{\alpha\ =\ x \ \mathrm{or}\ y}
\end{eqnarray}
where $e$ and $c$ are the electron charge and speed of light, respectively.
The two contributions take on well-known forms in the clean limit (Appendix \ref{app:A}), but are less familiar in the absence of translational symmetry  \cite{PhysRevB.75.024510,PhysRevB.47.7995}.
The diamagnetic response kernel is
\begin{eqnarray}
\label{supfl_dia}
K^{\rm{dia}}_{xx} = \dfrac{1}{N}\sum_{m,s} [\tilde{M}^{-1}]_{ms,ms}^{xx}\ f(E_{ms})
\end{eqnarray}
where $\tilde M^{-1}$ is the matrix representation of the inverse effective mass in the basis of eigenstates of the BdG Hamiltonian (Appendix \ref{app:B}) and $f(E_{ms})$ is the Fermi-Dirac function for the BdG energy eigenvalues.
\begin{figure}[tb]
    \centering
    \includegraphics[width=1\linewidth]{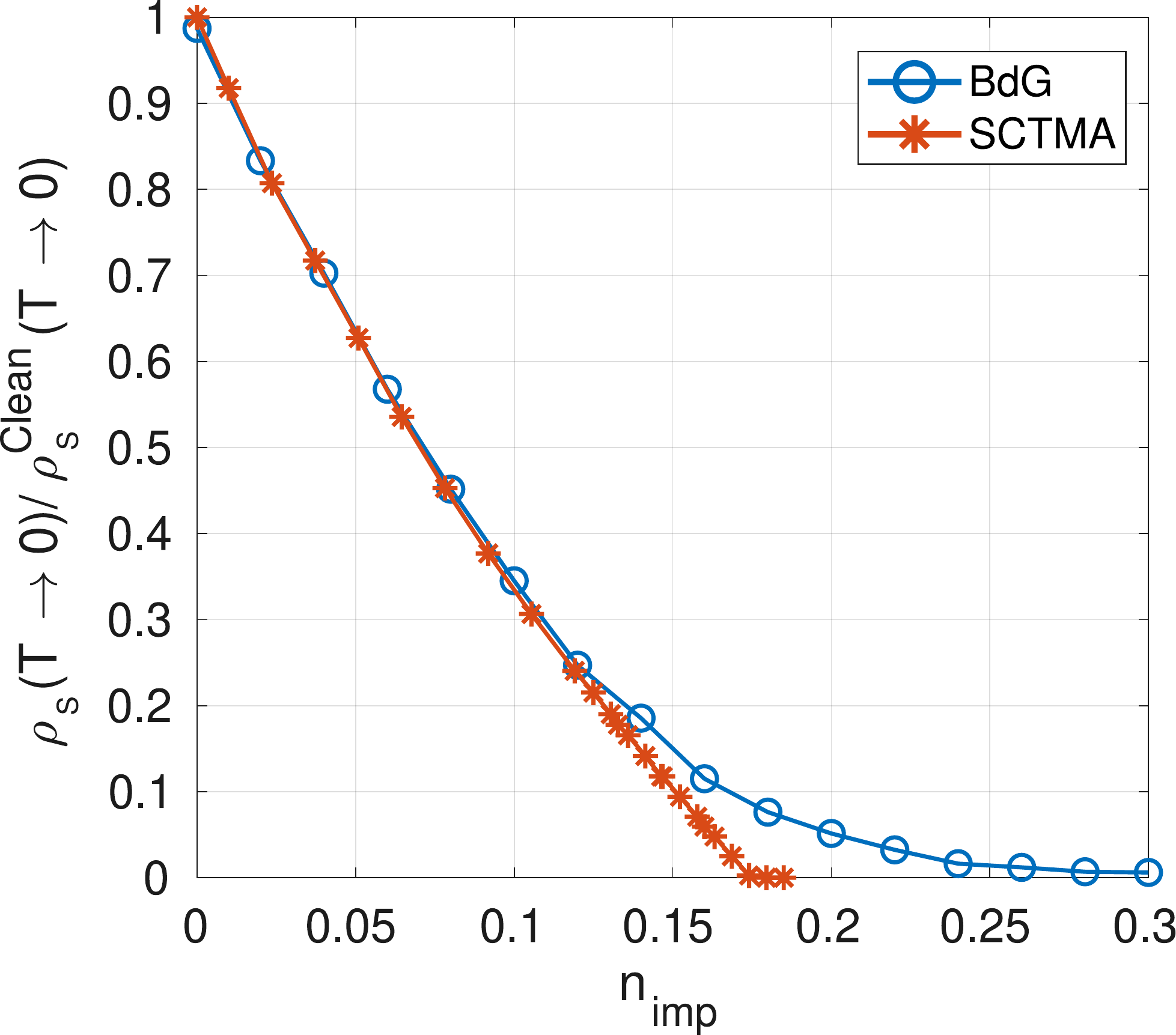}
    \caption{Superfluid density of the system in zero-temperature limit 
    averaged over $10$ random impurity configurations for the BdG calculation (blue) and SCTMA (red) as a function of impurity concentration.}
    \label{fig:superfluids_NOT_normalized_vs_imp_con}
\end{figure}
The paramagnetic response kernel is 
\begin{eqnarray}
\label{supfl_para}
K^{\rm{para}}_{xx} = 
\dfrac{1}{N}\sum_{m,n} \sum_{s,s^{\prime}} |\tilde{\zeta}_{ms,ns^{\prime}}|^2\ \dfrac{f(E_{ms}) - f(E_{ns^{\prime}})}{E_{ms} - E_{ns^{\prime}}}
\end{eqnarray}
Here, $\tilde{\zeta}$ is the representation of the quasiparticle velocity in the basis of eigenstates of the BdG Hamiltonian and, as above, $m,n$ and $s,s'$ are eigenenergy and pseudo-spin indices, respectively. $N$ is the number of lattice points. 
Detailed expressions for $\tilde M^{-1}$ and $\tilde \zeta$ are given in Appendix \ref{app:B}.

It is also usual to define a related quantity, the superfluid density, as
\begin{eqnarray}
\rho_{\mathrm{s}} = \dfrac{mc^2}{4\pi e^2} \lambda^{-2}.
\end{eqnarray}
{For a parabolic band, $m$ is the electron effective mass; otherwise, one may define $m$ from the clean-limit expression $\frac nm = K^\mathrm{dia}_{xx}(T=0)$, where $n$ is the electron density.
Then
\begin{equation}
    \rho_s(T) = n \cdot \frac{K^\mathrm{dia}_{xx}(T) + K^\mathrm{para}_{xx}(T)}{K^\mathrm{dia}_{xx}(0)|_\mathrm{clean}}.
\end{equation}
This choice correctly captures the fact that $\rho_s = n$ at $T=0$ in the clean limit, and is independent of either $T_c$ or the order parameter.}
For the qualitative behavior (and direct comparison with results in the supplementary information of Ref.~\cite{Li2021}), the fundamental constants have been suppressed in the plots for superfluid density. \subsubsection{Zero temperature}
  \begin{figure}
    \centering
    \includegraphics[width=\linewidth]{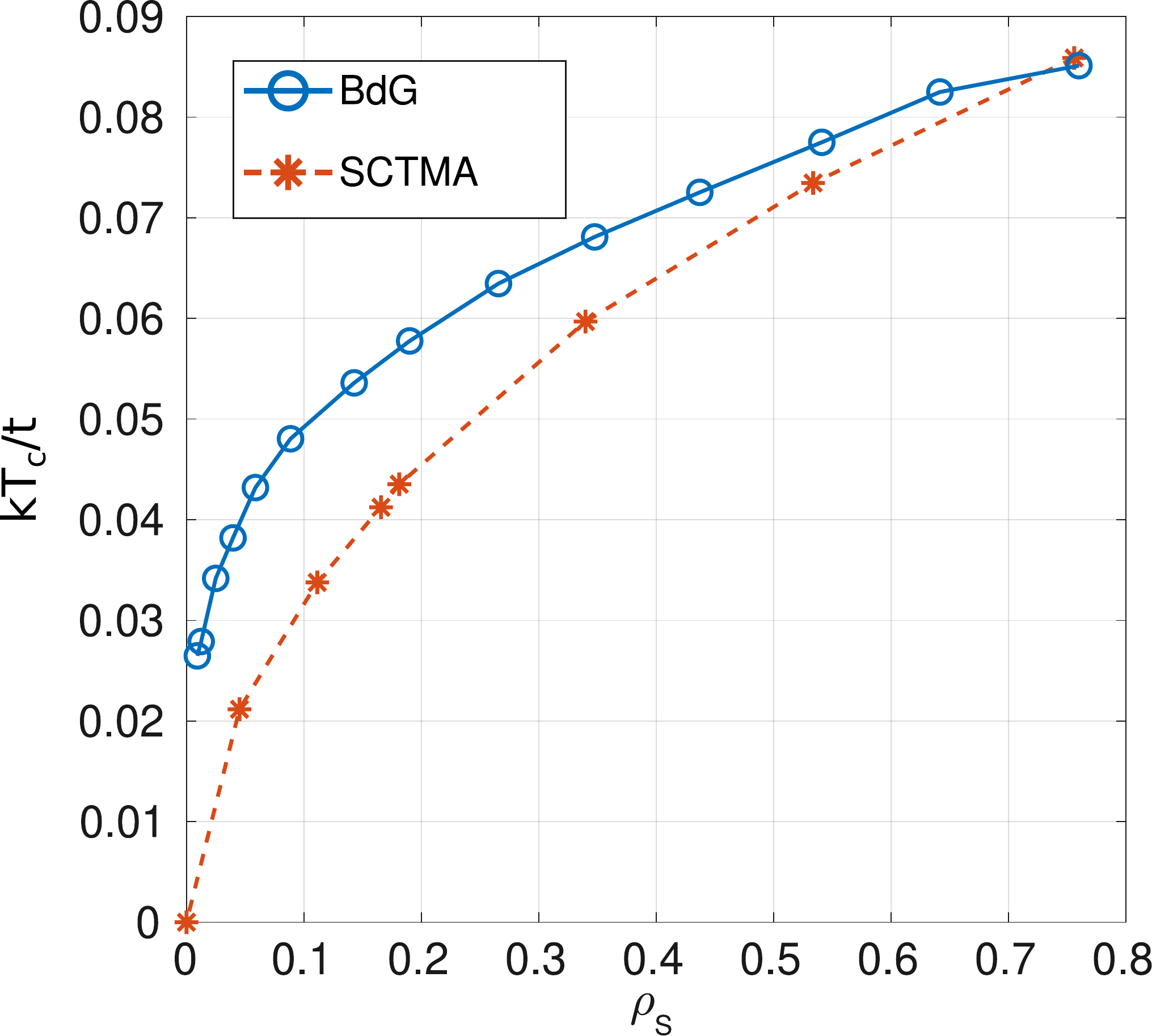}
    \caption{$T_c$ vs. superfluid density $\rho_s$ at zero $T$ for BdG (blue) and SCTMA (red).}
    \label{fig:rhos_vs_Tc}
\end{figure}
\begin{figure*}
    \centering
    \includegraphics[width=1\linewidth]{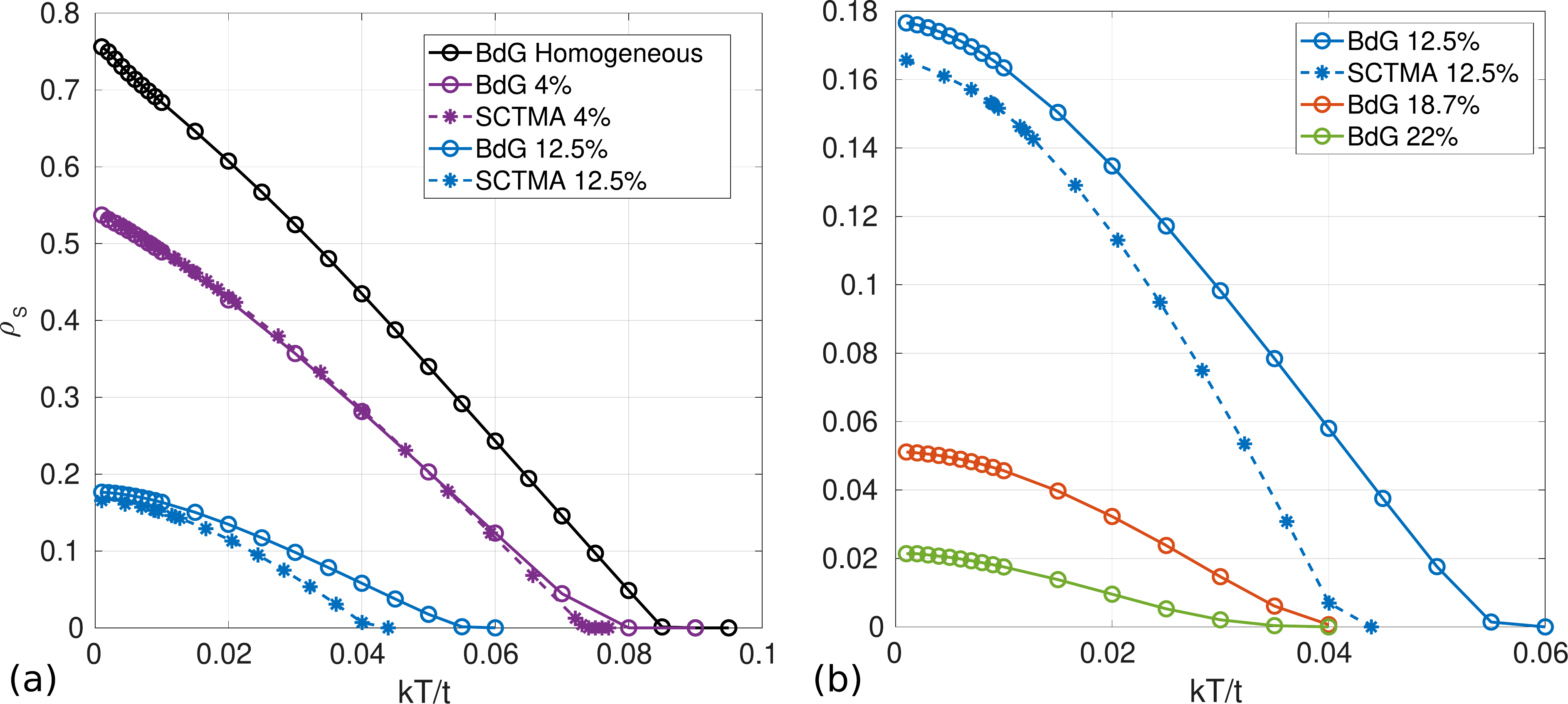}
    \caption{ Effect of disorder on the temperature-dependent 
    superfluid density.
    Superfluid densities at various temperatures were averaged over $40$ random impurity configurations (with 15$\times$15 supercells) at  impurity concentrations (a) 0\%, 4\%, and 12.5\%; (b) 12.5\%, 18.7\% and 22\%.   Note that, for the last two cases, SCTMA predicts that superconductivity is entirely supressed.
    }
    \label{fig:superfluids_normalized_12pt5_18pt7_22}
\end{figure*}
The zero-temperature superfluid density is a fundamental measure of the strength of superconductivity. In Fig. \ref{fig:superfluids_NOT_normalized_vs_imp_con}, we show calculations for $\rho_s$ from both our BdG numerical simulations and the SCTMA.  As expected, the superfluid density is seen to decrease monotonically as the impurity concentration is increased in both approximations. At lower impurity concentrations, the BdG and SCTMA predictions for the suppression of $\rho_s$ are virtually identical.  It is only when the superfluid density falls to about 10\% of its clean value that the two approaches differ significantly.  As is well-known from previous  work, $\rho_s$ vanishes linearly with disorder within the SCTMA, as in the classic Abrikosov-Gor'kov treatment of magnetic impurities in an $s$-wave superconductor\cite{Skalski1964}.    In the BdG case,  however, $\rho_s(0)$ has a concave-up tail that perisists to high impurity concentrations, similar to what was observed in
Ref.~\cite{Li2021}. This behavior is expected, because the SCTMA describes the effects of an average order parameter, whereas  in the BdG case, islands of nonzero $\rho_s$ in rare regions favorable for superconductivity\cite{Dodaro2018} persist well past the SCTMA critical concentration, and indeed well beyond the critical disorder where regions of nonzero order parameter percolate across the numerical sample (see Fig. \ref{fig:gapmaps}). 

It is also important to confirm the  ``non-BCS" quasiproportionality of $\rho_s(0)$ and $T_c$ reported in \cite{Bozovic2016} and reproduced in the subsequent SCTMA study\cite{PhysRevB.96.024501}.   In a clean BCS superconductor, the low-temperature superfluid density is equal to the density of conduction electrons in a normal metal.  When impurity scattering suppresses both $T_c$ and $\rho_s$ monotonically, a quasiproportionality between the two as observed in Ref. \cite{Bozovic2016} can result, however.    Accurate determination of $T_c$ is not trivial in a finite size system as considered in BdG.
Here to get each data point for $T_c$ in the figure, we take the superfluid density at three highest temperature points of the corresponding $\rho_{s}$ vs $T$ data and fit a quadratic curve through those three points to estimate the $T_c$. The result for   $T_c$ vs. $\rho_s(T \rightarrow 0)$ determined in this way is shown in Fig. \ref{fig:rhos_vs_Tc}, where we compare results for both approximations.  The functional form is clearly not consistent with the BCS result $\rho_s =n$, but neither is it consistent with the linear relation found in the experiment of Ref. \cite{Bozovic2016}.  We have verified that decreasing the impurity potential to correspond to the Born limit used in Ref. \cite{PhysRevB.96.024501} does not affect the SCTMA curve significantly.  The deviation from the quasilinear behavior is therefore due to band structure effects, specifically the proximity of the van Hove singularity to the Fermi level in the LSCO ARPES-derived band structure used in that work.
\subsubsection{Temperature dependence of superfluid density}
Next, we focus on the temperature dependence of the superfluid density in detail at a few specific impurity concentrations to see if the physics included in the BdG approach allows for an understanding of the approximate linear-$T$ behavior of the penetration depth reported in Ref. \cite{Bozovic2016}.  We were also motivated by the BdG calculations in Ref. \cite{Li2021}, which seemed to indicate that asymptotic linear-$T$  behavior was often found for quite disordered samples.

For this range of impurities, averaging was done over the $40\times 40$ system for forty random configurations at each value of impurity concentration and $15 \times 15$ supercells were used.
BdG results for the superfluid density temperature dependence, together with the equivalent result for the SCTMA, are presented in Fig. \ref{fig:superfluids_normalized_12pt5_18pt7_22}(a) and (b).  
The superfluid density in the clean, optimally doped cuprates typically decreases almost linearly as temperature increases\cite{Hardy1993}, reflecting the linear low energy  density of states of a clean $d$-wave superconductor.   However, strong disorder changes this linear-$T$ behavior into a quadratic behavior over an impurity bandwidth rather rapidly\cite{BonnHardy1993Zn}, as predicted by SCTMA\cite{prohammer1991,Hirschfeld:1993cka}.
Both approximations give very similar quantitative results over nearly the entire temperature range for the smaller disorder concentrations shown in Fig. \ref{fig:superfluids_normalized_12pt5_18pt7_22}(a), and the temperature dependences are qualitatively similar even for larger disorder concentrations, until the disorder concentration approaches the critical disorder for the destruction of superconductivity.   These discrepancies arise when the BdG superfluid density is suppressed to roughly 10\% of its clean value.   
  In overdoped LSCO, this would correspond to high levels of overdoping, near the end of the superconducting dome, e.g.\ for $T_c$ at or below 
$T_c=10$K \cite{Bozovic2016}.

In the weak scattering limit with generic bands, a larger concentration of impurities is necessary to create a clear quadratic low-$T$ behavior.  From the SCTMA viewpoint, in contrast to the strong impurity case, there is no analogous impurity bandwidth over which the $T^2$ is expected.  Indeed, Lee-Hone et al.\cite{PhysRevB.96.024501} found nearly linear-$T$ behavior over nearly the entire doping disorder range, but emphasized that a signifcant role was played by the particular Fermi surface of overdoped LSCO in the intermediate-to-low temperature regime.
\subsubsection{Supercell dependence of superfluid density}
\begin{figure}[tb]
    \centering
    \includegraphics[width=1\linewidth]{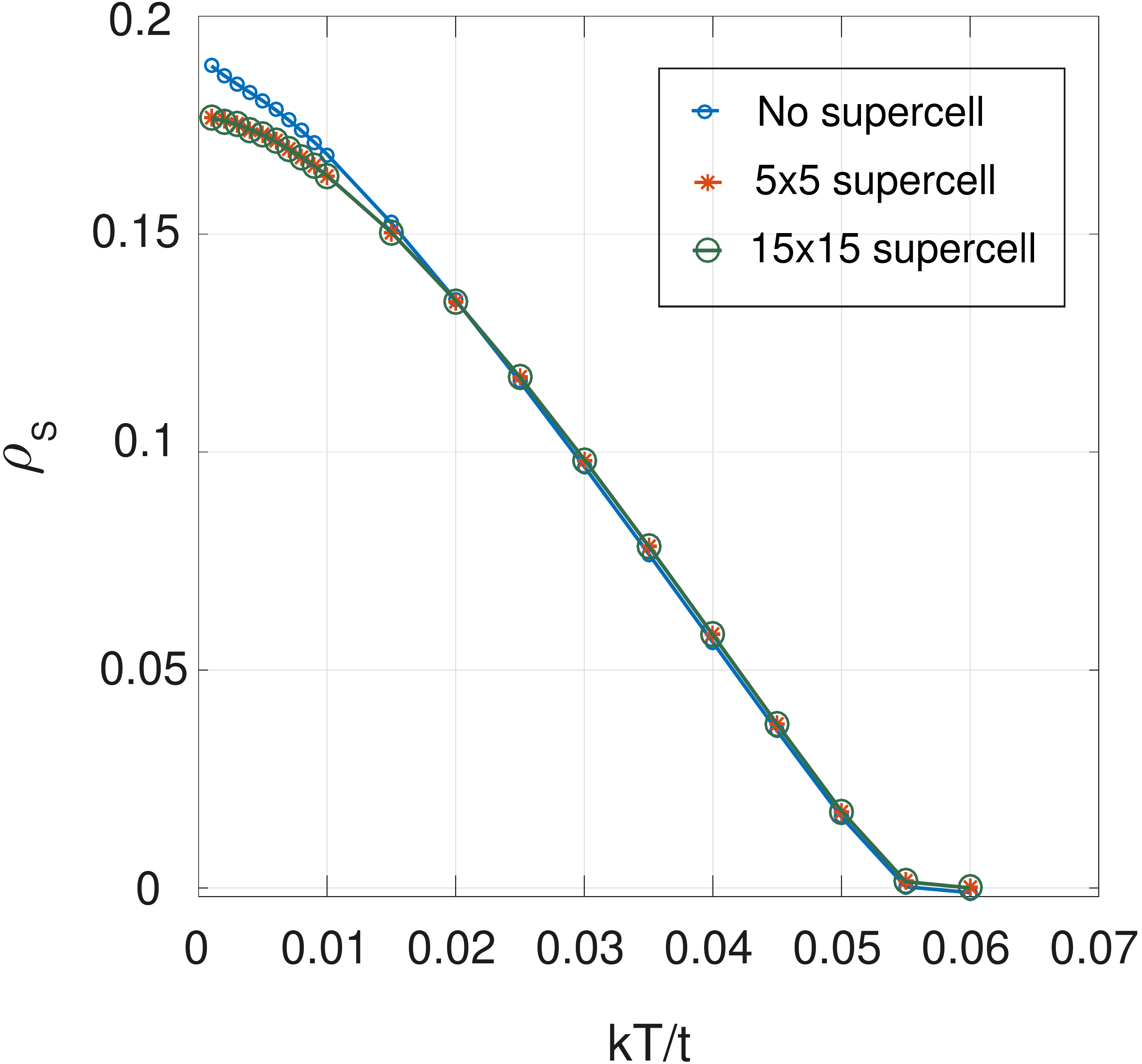}
    \caption[Supercell dependence of superfluid density]{{Supercell dependence of superfluid density.} Superfluid density averaged over 40 impurity configurations as a function of the temperature at fixed impurity concentration 12.5 \% with different number of supercells used. The superfluid density scale in this figure is not normalized with $\rho_{\mathrm{s}}({ T\rightarrow0})$ or $\rho^{\mathrm{Clean}}_{\mathrm{s}}({ T\rightarrow0})$   
}
\label{fig:superfluids_vs_temp_supcl_variation_12pt5}
\end{figure}

As mentioned above, some of the higher disorder concentrations of the $\rho_s$ vs. $T$ results shown in the supplementary information of  Ref. \cite{Li2021} appear to show low-$T$ linear behavior, in contrast to the results we have just presented.  Identical band structures and disorder potentials were used in both works.
To investigate possible origin of this mismatch, we studied the temperature variation of superfluid density with different number of supercells used at each of impurity concentrations $12.5\%$, $18.7\%$ and $22\%$.  The details of the implementation are given in Appendix A.
For  $18.7\%$ and $22\%$, we compared just between the effects of having no supercell and $15 \times 15$ supercells. For $12.5\%$ we additionally used a supercell lattice of intermediate size $5 \times 5$. The result is presented in Fig. \ref{fig:superfluids_vs_temp_supcl_variation_12pt5}.
It is to be emphasized that for small systems of finite size, the use of supercell lattice is quite important at low temperature to create a sufficient density of states  available within a narrow energy range. Remarkably, the low-temperature behavior of superfluid density without any supercell resembles the results from  \cite{Li2021} quite well, so we speculate that this effect is responsible for the discrepancy.  The deviation from increasingly more linear behavior becomes apparent in case of $18.7\%$ and $22\%$ impurities.

\begin{figure}[tb]
    \centering
    \includegraphics[width=1\linewidth]{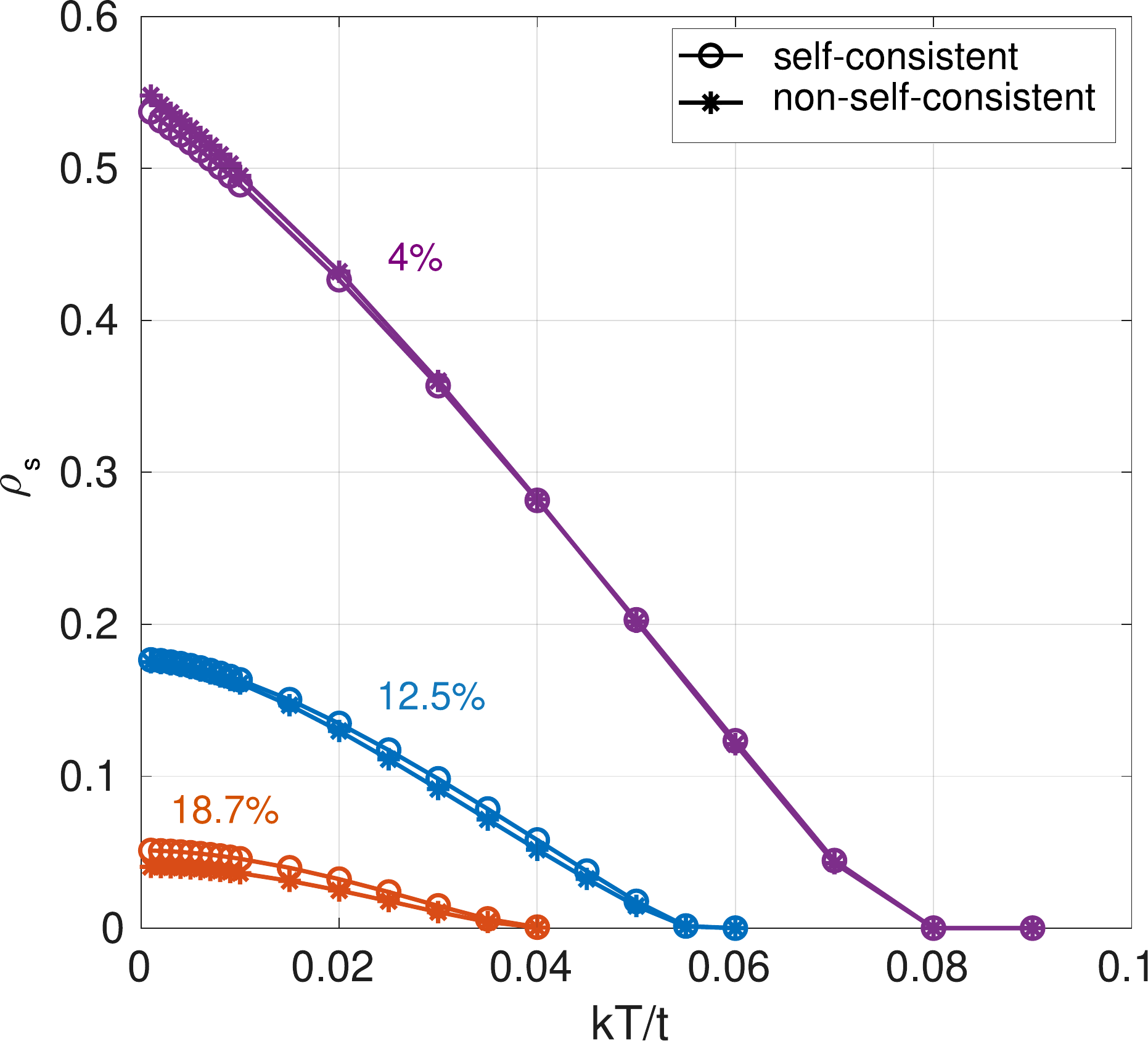}
    \caption[Superfluid density for self-consistent inhomogeneous superconducting order  and corresponding spatially averaged homogeneous (termed `non-self-consistent') superconducting order]{Superfluid density for self-consistent inhomogeneous superconducting order  and corresponding spatially averaged homogeneous (termed `non-self-consistent') superconducting order. The impurity concentration is $12.5\%$ and impurity configurations are same as in Fig.  \ref{fig:superfluids_vs_temp_supcl_variation_12pt5}. 
    }
    \label{fig:Non_self_const_vs_self_const_12pt5}
\end{figure}

\subsubsection{Superfluid density with self-consistent {vs. homogeneous average gap}}
There are several major differences between the BdG and SCTMA results presented here.  The BdG suffers from finite size corrections that need to be carefully estimated, as shown above; however it includes both the effects of self-consistent variation of the order parameter in response to a given disorder configuration, as well as quantum interference corrections (crossing diagrams in the self-energy and vertex corrections) neglected in the one-impurity T-matrix. 
The former source of difference is easy to check by performing non-self-consistent BdG calculations with artificial homogeneous order parameter.
For this, the inhomogeneous superconducting gap  for each configuration was replaced with its  average value at the corresponding temperature,
over the lattice before calculating the superfluid density. The result is presented in Fig. \ref{fig:Non_self_const_vs_self_const_12pt5}.
The self-consistent (inhomogeneous) result is seen to be smaller than the non-self-consistent (homogeneous average gap) result in the weakly disordered limit, where the superconductor is largely clean and the order parameter is suppressed only around individual impurities.  On the other hand, at large disorder, the system breaks up into distinct islands of relatively large superfluid density, such that the inhomogeneous result is slightly larger.   
 However, the differences between the homogeneous and inhomogeneous BdG results are always small.
This suggests that the effects of inhomogeneity are not important to understanding the differences between the two mean-field approximations presented here, BdG and SCTMA; rather, quantum interference diagrams beyond T-matrix neglected in the SCTMA drive the differences at larger disorder.    Homogeneity can of course become much more important to the {\it actual} superfluid density once phase fluctuations are included in the cases of very small superfluid density, however. These effects are beyond the scope of this work.
\section{Discussion}
In summary, we studied various superconducting properties in optimally doped to overdoped cuprates   using the impurity averaged self-consistent real-space BdG formalism with supercells and periodic boundary condition. The doping and impurity parameters { were taken to correspond exactly to the previous study \cite{Li2021} that unexpectedly found a linear $T$ dependence of the superfluid density for some highly disordered cases. The strength of the impurity  assumed in that work is probably significantly higher than realistic values for real spacer layer dopants \cite{https://doi.org/10.48550/arxiv.2206.01348}, and may be classified as moderate since the potential is a significant fraction of the bandwidth.}  Nevertheless we compared directly and found quadratic $T$-behavior for the analogous cases studied by Li et al.  
We have suggested here that the discrepancy lies in the use of to few states in the low energy region of the simulations of Ref. \cite{Li2021}.  We emphasize again, however, that these results were not the main point of their work, and our results confirm theirs in most other points of comparison.

We have shown that in this intermediate strength scattering regime, the quasiparticle density of states shows quite remarkably robust linear low-energy excitation spectrum above a disorder-dependent residual value {up to quite high disorder levels},  for both SCTMA and BdG calculations.   Nevertheless, the low- temperature superfluid density behavior apparently displays the usual quadratic low-$T$ temperature dependence expected for a dirty $d$-wave superconductor.  This is an unexpected result based on the usual BCS argument, whereby the leading $\omega$ term in $\rho(\omega)$ near the Fermi level determines the low temperature $T$ dependence of the superfluid density.  Possible sources of this discrepancy include Andreev scattering from gap inhomogeneities, and quantum interference from multi-impurity scattering (crossing diagrams).  Fig.~\ref{fig:Non_self_const_vs_self_const_12pt5} shows that gap inhomogeneities do not significantly change the low-$T$ superfluid density.  We therefore conclude that the multi-impurity scattering effects are responsible for the deviation from the BCS relation between the density of states $\rho(\omega)$ and the superfluid density $\rho_s(T)$.

Gap inhomogeneity causes shifts in $\rho_s$ at low $T$.  The direction of the shifts is seen in Fig. \ref{fig:Non_self_const_vs_self_const_12pt5} to depend on the impurity concentration.  For low impurity concentration, the gap inhomogeneities are isolated, and therefore lead to a small local suppression of $\rho_s$ near each impurity. This adds an Andreev scattering channel that further suppresses the superfluid density.  On the other hand,  for large impurity concentrations, the inhomogeneity increases $\rho_s$ slightly since now large puddles of superconductor preferentially form in cleaner regions where ordinary impurity scattering does not suppress $\rho_s$. 

These studies suggest that the self-consistent T-matrix approximation  works very well compared to full quantum calculations for weak to intermediate strength scattering potentials, even up to quite large disorder concentrations, in contrast to strong scattering potentials\cite{Atkinson2000}.  Despite the fact 
that we found here results for the superfluid density displaying substantial quadratic behavior at low $T$, as expected for SCTMA at high concentrations even for weak scatterers, it does not invalidate the result of Ref. \cite{PhysRevB.96.024501}.  This is because unlike this reference, our studies neglect the effect of the doping on the van Hove singularity near the Fermi level, as occurs in  LSCO.

\textit{Acknowledgements:} We thank B.M. Andersen, C. N. Brei{\o}, D.M. Broun, S.A. Kivelson and M.A. Sulangi  for useful discussions.  P.J.H. and M.P. acknowledge support from NSF-DMR-1849751.  W.A.A. acknowledges support from the Natural Sciences and
Engineering Research Council (NSERC) of Canada. A.K. acknowledges support by the Danish National Committee for Research Infrastructure (NUFI) through the ESS-Lighthouse Q-MAT.

\appendix
\section{Supercell Method}
\label{supercell_method}

The numerical calculations in this article are based on the Bogoliubov-de Gennes (BdG) method. For a homogeneous system, results from the real space BdG calculations can also be readily obtained from the corresponding momentum space equations because of translation invariance. One can freely choose as dense a momentum space grid as one wants which in real space would correspond to a bigger and bigger part of an infinite homogeneous lattice. This larger portion of lattice, however, can also be obtained from stitching the `lattice-portion corresponding to the coarse momentum grid' together one after another in one or more of the spatial directions. The `entire lattice-portion corresponding to the coarse momentum grid' can be called a supercell in real space and the momentum corresponding to the periodicity of these supercell is called supercell momentum. Obviously, the neighboring supercells couple with each other through the sites near their margin/boundary only, the supercell being just an imaginary construct of grouping lattice sites together. Hence, for a square lattice (which has been used in this article), only nearest neighbor (NN) and next nearest neighbor (NNN) supercells can couple with each other unless the supercell size is not critically small compared to the hopping range of the original lattice sites. 
  
The corresponding scenario of stitching together the inhomogeneous lattices of a finite size is particularly useful for increasing smoothness of the spectrum. In this case, within the supercell there is no notion of momentum because of lack of translational invariance and the eigenstates of the inhomogeneous finite lattice are to be obtained by numerical diagonalization, often starting from the real space basis. However, subsequently, the use of many supercells effectively creates a narrow band of energy around each of these numerically obtained eigenenergies, thereby increasing the spectrum resolution. 

\begin{figure}[tb]
    \centering
    \includegraphics[width=1\linewidth]{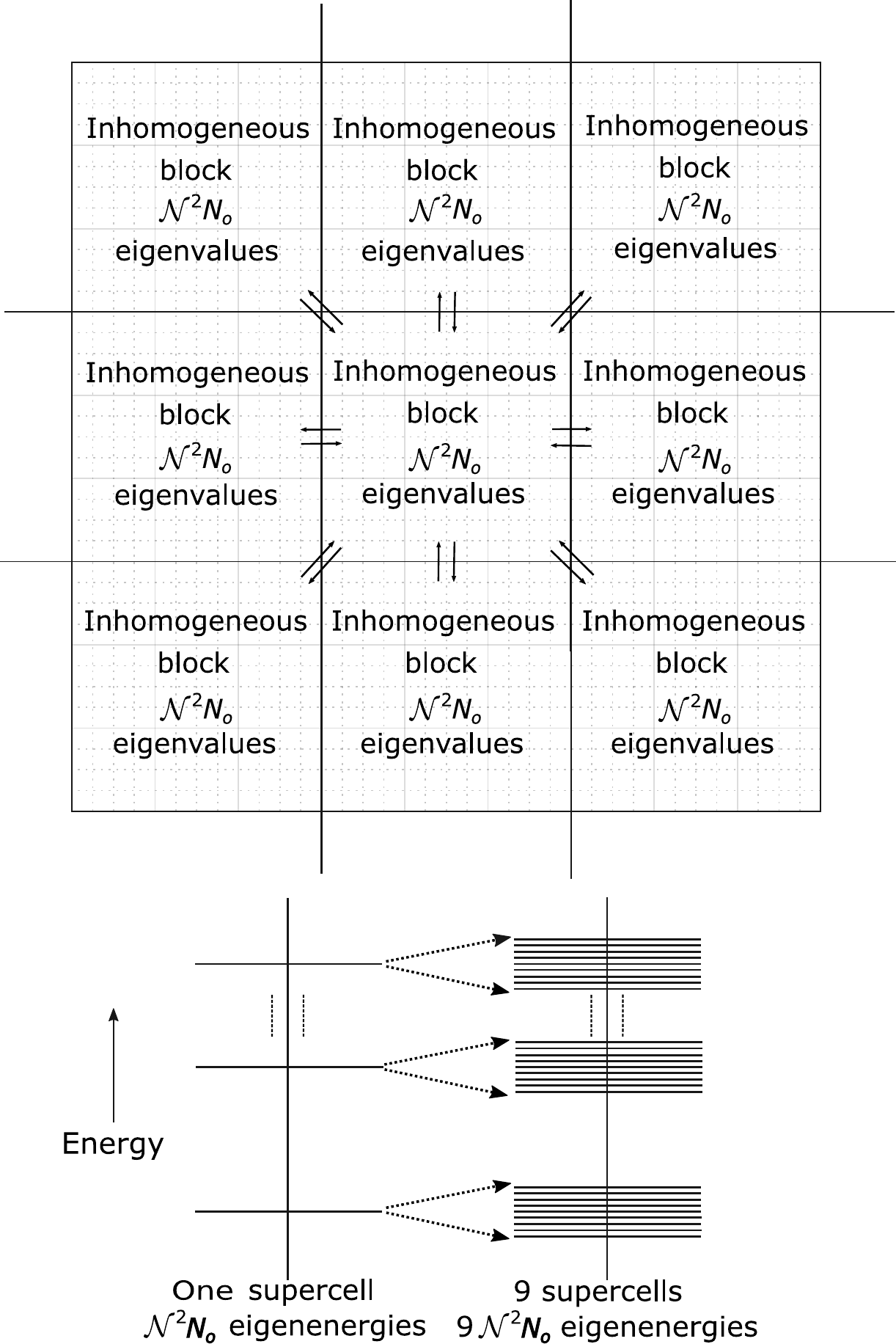}
    \caption[Illustration of the supercell method]{{Illustration of the supercell method:} A supercell lattice of size $3 \times 3$  (periodically continued both horizontally and vertically) consisting of supercells of size $\mathcal{N} \times \mathcal{N}$ where each lattice site hosts $1$ orbital. The double-way arrows show the supercell-supercell coupling in which the sites near the supercell boundaries participate.}
    \label{fig:supercell}
\end{figure}

If $M \times M$ supercells are used, the supercell BdG equation in matrix form can be written as
\begin{equation}
    \begin{pmatrix}
    h(\mathbf{K}) & \Delta(\mathbf{K}) \\
    \Delta^{\dagger}(\mathbf{K}) & -h^T(-\mathbf{K})
    \end{pmatrix}
    %\begin{pmatrix}
    %U_{ns}(\mathbf{K}) \\
    U_{:,ns}(\mathbf{K})
    %\end{pmatrix}
    =
    E_{ns}(\mathbf{K})
        U_{:,ns}(\mathbf{K})
    %\begin{pmatrix}
    %U_{n\Uparrow}(\mathbf{K}) \\
    %U_{n\Downarrow}(\mathbf{K})
    %\end{pmatrix}
    \label{A1}
\end{equation}
with 
\begin{align}
  h(\mathbf{K}) = \sum_{I} T_{I0} e^{i\mathbf{K}\cdot\mathbf{R}_I}, \label{A2}
  \\
  \Delta(\mathbf{K}) = \sum_{I} \Delta_{I0} e^{i\mathbf{K}\cdot\mathbf{R}_I}.
\end{align}
Here $\mathbf{K}$ is supercell momentum, $\mathbf{R}_{I}$ is the coordinate of the $I$th supercell, $T_{I0}$ is the supercell hopping matrix (of size $\mathcal{N}^2N_o \times \mathcal{N}^2N_o$ where the size of the supercell is $\mathcal{N} \times \mathcal{N}$ and each site has $N_o$ orbitals) between the supercell at origin and the $I$th supercell, and  $\Delta_{I0}$ is the supercell gap matrix (of size $\mathcal{N}^2N_o \times \mathcal{N}^2N_o$) between the supercell at origin and the $I$th supercell.
The matrix $U$ diagonalizes the Hamiltonian and $ U_{:,ns}(\mathbf{K})$ is the $ns$ column-vector of $U(\mathbf{K})$.
In this work we have $N_o=1$ and number of lattice points $N=\mathcal{N}^2$. Most of the matrix elements of $T_{I0}$ and $\Delta_{I0}$ (for $I\ne 0$) are zero unless they involve sites near the supercell-boundary which maintain the supercell-supercell coupling. $T_{00}$ and $\Delta_{00}$ are exactly same as the hopping matrix and the gap matrix of the $\mathcal{N} \times \mathcal{N}$ inhomogeneous lattice, solved by explicit iterative self-consistent diagonalization to get eigenenergies indexed by $n$. $\mathbf{K}$ can take values as $K_x,K_y=\dfrac{m}{M},\ m=0,1,2,...M-1$ in units of $2\pi/\mathcal{N}a$ where $a$ is lattice constant, and for each  $\mathbf{K}$ the supercell BdG diagonalization is done just once to obtain the eigenvectors $U_{:,ns}(\mathbf{K})$, each of size $2\mathcal{N}^2N_o$. So, as already mentioned, corresponding to each $n$, a band of additional $M^2 - 1$ eigenenergies are obtained. The picture is schematically presented in Fig. \ref{fig:supercell}
\section{Clean penetration depth in real and $k$ space}
\label{app:A}
It is useful for comparison to real-space BdG and SCTMA calculations to have full  results for the superfluid density of the corresponding clean system, which can be calculated more accurately in momentum space for bands $\xi_{\bf k}$ and gap function $\Delta_{\bf k}$ as given in the main text.  The result is
(see, e.g. \cite{PhysRevB.70.054510,PhysRevB.47.7995}) 
% \begin{widetext}

\begin{equation}
\begin{split}
\lambda^{-2}_{xx} &= \dfrac{4\pi e^{2}}{c^{2}}\Bigg[\sum_{\textbf{k}} \dfrac{\partial^{2}\xi_{\textbf{k}}}{\partial k_{x}^2}\Bigg(1 - \dfrac{\xi_{\textbf{k}}}{E_{\textbf{k}}} \tanh{\dfrac{\beta E_{\textbf{k}}}{2}}\Bigg)
\\ &-\dfrac{\beta }{2}\sum_{\textbf{k}}\Bigg(\dfrac{\partial\xi_{\textbf{k}}}{\partial k_{x}}\Bigg)^2\mathrm{sech}^{2}\dfrac{\beta E_{\textbf{k}}}{2}\Bigg]{
}
\end{split}
\label{eq:rhos_clean_v1}
\end{equation}
% \end{widetext}
where $\xi_\textbf{k}$ is the single particle energy with respect to chemical potential as specified in the manuscript, $E_\textbf{k} = \sqrt{\xi_\textbf{k}^2 + \Delta_\textbf{k}^2}$.
The first sum gives the diamagnetic contribution and the second sum is the paramagnetic contribution.
We have tested that the calculated penetration depth using this formulae is quite strongly dependent on the number of momentum space points. The diamagnetic term has contributions from the full Brillouin zone and the paramagnetic term has dominant contributions from the nodal areas. The diamagnetic term can be written in a different form using integration by parts, which leads to an alternate expression for the penetration depth \cite{PhysRevB.70.054510}
% \begin{widetext}
\begin{equation}
\begin{split}
\lambda^{-2}_{xx} &= \dfrac{4\pi e^{2}}{c^{2}}\sum_{\textbf{k}}
\Bigg[\Bigg(\dfrac{\partial\xi_{\textbf{k}}}{\partial k_{x}}\Bigg)^2\Bigg(\dfrac{\Delta_\textbf{k}}{E_\textbf{k}}\Bigg)^2 
\\&-\Bigg(\dfrac{\partial\xi_{\textbf{k}}}{\partial k_{x}}\Bigg) \Bigg(\dfrac{\partial\Delta_{\textbf{k}}}{\partial k_{x}}\Bigg)\dfrac{\Delta_\textbf{k}\xi_\textbf{k}}{E_\textbf{k}^2}\Bigg] 
% \\&\ \ \ \ 
\Bigg[\dfrac{1}{E_\textbf{k}} - \dfrac{\partial}{\partial E_\textbf{k}}\Bigg]\tanh{\dfrac{\beta E_\textbf{k}}{2}}
\end{split}
\end{equation}
% \end{widetext}
Although mathematically equivalent to Eq.~(\ref{eq:rhos_clean_v1}), this alternate expression converges differently as a function of the number of $k$ points in the calculation. For finite systems, this can lead to inaccurate results for the superfluid density particularly at low temperatures. In the real space calculation, we are using the (effective) multi-band analog of Eq. (\ref{eq:rhos_clean_v1}) which shows a dependence on the number of supercell $k$-points especially at low temperatures, as demonstrated in Fig. \ref{fig:superfluids_vs_temp_supcl_variation_12pt5}.

\section{Inverse mass and velocity}
\label{app:B}

The Hamiltonian for an $N$ site system in the usual fermionic basis is
\begin{equation}
  \hat H = c^{\dagger} H c
   \ \ \ \ \ \mathrm{with} \ \ \ \ \ 
  H = 
\begin{pmatrix}
h & \Delta \\
\Delta^{\dagger} & -h^T
\end{pmatrix}, 
\label{C1}
\end{equation}
where $H$ is a $2N\times2N$ matrix, $h$ and $\Delta$ are $N\times N$ matrices with matrix elements
\begin{eqnarray}
h_{ij} &=& t_{ij} + (w_i - \mu) \delta_{ij} \\
\Delta_{ij}&=& V\langle c_{i\uparrow}c_{j\downarrow} \rangle,
\end{eqnarray}
and $c$ and $c^{\dagger}$ are respectively column and row vectors defined as 
\begin{eqnarray}
c^{\dagger} = 
\begin{pmatrix}
c^{\dagger}_{1\uparrow} & ... & c^{\dagger}_{N \uparrow}
& c_{1\downarrow} & ... & c_{N \downarrow}
\end{pmatrix} 
.
\end{eqnarray}

$H$ is diagonalized by a unitary matrix $U$ such that
% \begin{eqnarray}
    $U^{\dagger} H U = E$,
% \end{eqnarray}
with $E$ a diagonal matrix of eigenenergies.
The quasiparticle operators for the diagonalized Hamiltonian are  
\begin{align}
\Gamma^{\dagger} = c^{\dagger} U = 
\begin{pmatrix}
\gamma^{\dagger}_{1\Uparrow} & ... & \gamma^{\dagger}_{N \Uparrow}
& \gamma^{\dagger}_{1\Downarrow} & ... & \gamma^{\dagger}_{N \Downarrow}
\end{pmatrix}
\end{align}
where in $\gamma^\dagger_{n s}$, $n$ labels an eigenstate and $s$ a pseudo-spin. The eigenenergies satisfy $E_{n \Uparrow} >0$ and  $E_{n \Downarrow} < 0$.

The paramagnetic part of the electromagnetic response kernel 
%related to the inverse squared penetration depth 
is given by \cite{PhysRevB.75.024510}
\begin{align}
K^{\rm{para}}_{xx} = \frac{1}{N}
\sum_{m,n} \sum_{s,s^{\prime}} |\tilde{\zeta}_{ms,ns^{\prime}}|^2\ \dfrac{f(E_{ms}) - f(E_{ns^{\prime}})}{E_{ms} - E_{ns^{\prime}}}
\label{C6}
\end{align}
where $m,n$ are eigenenergy indices, $s,s^{\prime}$ are pseudo-spin indices and
\begin{eqnarray}
\tilde{\zeta} = U^{\dagger}
\begin{pmatrix}
\zeta & 0 \\
0 & \zeta
\end{pmatrix}
U
\end{eqnarray}
is the representation of the electron velocity operator $\zeta$ in the basis of eigenstates of the Hamiltonian.  In the basis of lattice sites, $\zeta$ is an $N \times N$ matrix with matrix elements 
$\zeta_{lm} = i(x_l - x_m) t_{lm}$, where $t_{lm}$ is the tightbinding hopping matrix element between sites $l$ and $m$, and $x_l$ and $x_m$ are the x-coordinates of the lattice site location.  This form of the velocity operator 
originates in linear response theory from the Peierls substitution in the presence of a perturbing field.

The diamagnetic part of the electromagnetic response kernel is
\begin{eqnarray}
K^{\rm{dia}}_{xx} = \frac{1}{N} \sum_{m,s} (\tilde{M}^{-1})_{ms,ms}^{xx}\ f(E_{m{ s}}),
\label{C8}
\end{eqnarray}
with 
\begin{eqnarray}
\tilde{M}^{-1} = U^{\dagger} 
\begin{pmatrix}
M^{-1} & 0 \\
0 & - M^{-1}
\end{pmatrix}
U,
\end{eqnarray}
and
\begin{eqnarray}
(M^{-1})_{ij} = -t_{ij} (x_i - x_j)^2.
\label{C10}
\end{eqnarray}
The inverse squared penetration depth and superfluid density are then given by
\begin{eqnarray}
\lambda^{-2}_{xx} = 
\dfrac{4\pi e^{2}}{c^2} 
%\langle 
\left ( K^{\rm{dia}}_{xx} + K^{\rm{para}}_{xx} \right ),
%\rangle_{\alpha\ =\ x \ \mathrm{or}\ y}
\end{eqnarray}
and
\begin{eqnarray}
\rho_{\mathrm{s}} = \dfrac{mc^2}{4\pi e^2} \lambda^{-2}_{xx}.
\end{eqnarray}
The formalism described in this section can be extended to the supercell method in order to facilitate numerical calculations. To this end, one needs to Fourier transform the Hamiltonian Eq.~(\ref{C1}) to a form like that in Eq.~(\ref{A1}); the velocity and mass operators can be Fourier transformed using the prescription in Eq.~(\ref{A2}) and acquire phase factors for displacements $x_i-x_j$ connecting different supercells. The sums over bands in Eqs.~(\ref{C6}) and (\ref{C8}) then pick up an additional sum over  supercell momenta $\mathbf K$.

\section{Dirty $d$-wave theory}
\label{app:C}
In the presence of impurities, the superconducting Green's function is 
\begin{equation}
G_{\bf k}(i\omega_n) = [i\omega_n \tau_0 - \xi_{\bf k} \tau_3 - \Delta_{\bf k} \tau_1 - \Sigma(i\omega_n)]^{-1},
\end{equation}
where $\Sigma(i\omega_n)$ is a $2\times 2$ matrix that incorporates the effects of impurity scattering.  Within SCTMA, the self-energy for pointlike impurities is obtained from the T-matrix via
\begin{equation}
\Sigma(i\omega_n) = n_\mathrm{imp} T(i\omega_n)
\end{equation}
where the $2\times 2$ T-matrix is 
\begin{equation}
T = [\tau_0 V^{-1}-\tau_3 g(i\omega_n)]^{-1} \tau_3
\end{equation}
with 
\begin{equation}
g(i\omega_n) = \frac{1}{N} \sum_{\bf k} G_{{\bf k}}(i\omega_n).
\end{equation}
The impurity potential is set to $V=1$.

For a $d$-wave gap with bond order parameters $\Delta_{\pm x} = -\Delta_{\pm y}=\Delta_0/2$, we have $\Delta_{{\bf k}} = \Delta_0 \eta_{\bf k}$ and
\begin{equation}
\Delta_0 = -\frac{J}{N}\sum_{\bf k} \eta_{\bf k} \langle c_{-{\bf k}\downarrow} c_{{\bf k} \uparrow} \rangle = - \frac J{\beta N}\sum_{n, {\bf k}} \eta_{{\bf k}}  [G_{{\bf k}}(i\omega_n)]_{12},
\end{equation}
 with $\eta_{\bf k} = \cos k_x - \cos k_y$.  It is common to write
 \begin{eqnarray}
 \tilde \omega_n &=& \omega_n - \frac 12 \mbox{Im }\mbox{Tr }{[\Sigma (i\omega_n) \tau_0]} \\
 \tilde \xi_{\bf k}(i\omega_n) &=& \xi_{\bf k} + \frac 12 \mbox{Re }\mbox{Tr }{[\Sigma (i\omega_n) \tau_3]} \\
 \tilde \Delta_{\bf k}(i\omega_n) &=& \Delta_{\bf k} + \frac 12 \mbox{Re }\mbox{Tr }{[\Sigma (i\omega_n) \tau_1]} 
 \end{eqnarray}
 so that 
 \begin{equation}
\Delta_0 = - \frac J{\beta N}\sum_{n, {\bf k}} \eta_{{\bf k}}  \frac{\tilde \Delta_{\bf k}}{(i\tilde \omega_n)^2 - \tilde \xi_{\bf k}^2 - \tilde \Delta_{\bf k}^2}
\end{equation}
This is further simplified for pointlike impurities and $d$-wave superconductivity, for which the off-diagonal matrix elements of the self-energy vanish and $\tilde \Delta_{\bf k}=\Delta_{\bf k}$.  Because the gap is unrenormalized, Anderson's theorem breaks down and the superconducting nonmagnetic impurities reduce the order parameter through pairbreaking.

It is common to ignore the chemical potential shift that comes from the real part of the diagonal elements of $\Sigma(i\omega_n)$.  If one makes this assumption, then the dispersion $\xi_{\bf k}$ is also unrenormalized; that is, 
$\tilde \xi_{\bf k} = \xi_{\bf k}$.  We do not make this assumption here because we wish to include the chemical potential shift that comes from impurity doping.  We have checked that the  shifts in electron density so obtained are the same as in the BdG calculations.

With these approximations, the diamagnetic kernel is 
\begin{eqnarray}
{K}^\mathrm{dia}_{xx} &=& \frac{1}{N\beta } \sum_{{\bf k},n} \frac{\partial^2 \xi_{\bf k}}{\partial k_x^2} \mathrm{Tr}\left [ G_{\bf k}(i\omega_n) \tau_3 \right] 
\end{eqnarray}
and the paramagnetic kernel is
\begin{equation}
{K}^\mathrm{para}_{xx} = \frac{1}{N\beta} 
\sum_{{\bf k}, n} \left( \frac{\partial \xi_{\bf k}}{\partial k_x} \right )^2 \mathrm{Tr}\left[ G_{{\bf k}}(i\omega_n) G_{{\bf k}}(i\omega_n) \right ].
\end{equation}

\bibliography{Thesis_references.bib}

\begin{thebibliography}{30}%
\makeatletter
\providecommand \@ifxundefined [1]{%
 \@ifx{#1\undefined}
}%
\providecommand \@ifnum [1]{%
 \ifnum #1\expandafter \@firstoftwo
 \else \expandafter \@secondoftwo
 \fi
}%
\providecommand \@ifx [1]{%
 \ifx #1\expandafter \@firstoftwo
 \else \expandafter \@secondoftwo
 \fi
}%
\providecommand \natexlab [1]{#1}%
\providecommand \enquote  [1]{``#1''}%
\providecommand \bibnamefont  [1]{#1}%
\providecommand \bibfnamefont [1]{#1}%
\providecommand \citenamefont [1]{#1}%
\providecommand \href@noop [0]{\@secondoftwo}%
\providecommand \href [0]{\begingroup \@sanitize@url \@href}%
\providecommand \@href[1]{\@@startlink{#1}\@@href}%
\providecommand \@@href[1]{\endgroup#1\@@endlink}%
\providecommand \@sanitize@url [0]{\catcode `\\12\catcode `\$12\catcode
  `\&12\catcode `\#12\catcode `\^12\catcode `\_12\catcode `\%12\relax}%
\providecommand \@@startlink[1]{}%
\providecommand \@@endlink[0]{}%
\providecommand \url  [0]{\begingroup\@sanitize@url \@url }%
\providecommand \@url [1]{\endgroup\@href {#1}{\urlprefix }}%
\providecommand \urlprefix  [0]{URL }%
\providecommand \Eprint [0]{\href }%
\providecommand \doibase [0]{https://doi.org/}%
\providecommand \selectlanguage [0]{\@gobble}%
\providecommand \bibinfo  [0]{\@secondoftwo}%
\providecommand \bibfield  [0]{\@secondoftwo}%
\providecommand \translation [1]{[#1]}%
\providecommand \BibitemOpen [0]{}%
\providecommand \bibitemStop [0]{}%
\providecommand \bibitemNoStop [0]{.\EOS\space}%
\providecommand \EOS [0]{\spacefactor3000\relax}%
\providecommand \BibitemShut  [1]{\csname bibitem#1\endcsname}%
\let\auto@bib@innerbib\@empty
%</preamble>
\bibitem [{\citenamefont {Bo{\v{z}}ovi{\'{c}}}\ \emph
  {et~al.}(2016)\citenamefont {Bo{\v{z}}ovi{\'{c}}}, \citenamefont {He},
  \citenamefont {Wu},\ and\ \citenamefont {Bollinger}}]{Bozovic2016}%
  \BibitemOpen
  \bibfield  {author} {\bibinfo {author} {\bibfnamefont {I.}~\bibnamefont
  {Bo{\v{z}}ovi{\'{c}}}}, \bibinfo {author} {\bibfnamefont {X.}~\bibnamefont
  {He}}, \bibinfo {author} {\bibfnamefont {J.}~\bibnamefont {Wu}},\ and\
  \bibinfo {author} {\bibfnamefont {A.~T.}\ \bibnamefont {Bollinger}},\
  }\bibfield  {title} {\bibinfo {title} {Dependence of the critical temperature
  in overdoped copper oxides on superfluid density},\ }\href
  {https://doi.org/10.1038/nature19061} {\bibfield  {journal} {\bibinfo
  {journal} {Nature}\ }\textbf {\bibinfo {volume} {536}},\ \bibinfo {pages}
  {309} (\bibinfo {year} {2016})}\BibitemShut {NoStop}%
\bibitem [{\citenamefont {{Mahmood}}\ \emph {et~al.}(2019)\citenamefont
  {{Mahmood}}, \citenamefont {{He}}, \citenamefont {{Bo{\v z}ovi{\'c}}},\ and\
  \citenamefont {{Armitage}}}]{Mahmood2019}%
  \BibitemOpen
  \bibfield  {author} {\bibinfo {author} {\bibfnamefont {F.}~\bibnamefont
  {{Mahmood}}}, \bibinfo {author} {\bibfnamefont {X.}~\bibnamefont {{He}}},
  \bibinfo {author} {\bibfnamefont {I.}~\bibnamefont {{Bo{\v z}ovi{\'c}}}},\
  and\ \bibinfo {author} {\bibfnamefont {N.~P.}\ \bibnamefont {{Armitage}}},\
  }\bibfield  {title} {\bibinfo {title} {{Locating the missing superconducting
  electrons in overdoped cuprates}},\ }\href
  {https://doi.org/10.1103/PhysRevLett.122.027003} {\bibfield  {journal}
  {\bibinfo  {journal} {Phys. Rev. Lett.}\ }\textbf {\bibinfo {volume} {122}},\
  \bibinfo {pages} {027003} (\bibinfo {year} {2019})}\BibitemShut {NoStop}%
\bibitem [{\citenamefont {Li}\ \emph {et~al.}(2021)\citenamefont {Li},
  \citenamefont {Kivelson},\ and\ \citenamefont {Lee}}]{Li2021}%
  \BibitemOpen
  \bibfield  {author} {\bibinfo {author} {\bibfnamefont {Z.-X.}\ \bibnamefont
  {Li}}, \bibinfo {author} {\bibfnamefont {S.~A.}\ \bibnamefont {Kivelson}},\
  and\ \bibinfo {author} {\bibfnamefont {D.-H.}\ \bibnamefont {Lee}},\
  }\bibfield  {title} {\bibinfo {title} {Superconductor-to-metal transition in
  overdoped cuprates},\ }\href {https://doi.org/10.1038/s41535-021-00335-4}
  {\bibfield  {journal} {\bibinfo  {journal} {npj Quantum Materials}\ }\textbf
  {\bibinfo {volume} {6}},\ \bibinfo {pages} {36} (\bibinfo {year}
  {2021})}\BibitemShut {NoStop}%
\bibitem [{\citenamefont {Kogan}(2013)}]{PhysRevB.87.220507}%
  \BibitemOpen
  \bibfield  {author} {\bibinfo {author} {\bibfnamefont {V.~G.}\ \bibnamefont
  {Kogan}},\ }\bibfield  {title} {\bibinfo {title} {Homes scaling and bcs},\
  }\href {https://doi.org/10.1103/PhysRevB.87.220507} {\bibfield  {journal}
  {\bibinfo  {journal} {Phys. Rev. B}\ }\textbf {\bibinfo {volume} {87}},\
  \bibinfo {pages} {220507} (\bibinfo {year} {2013})}\BibitemShut {NoStop}%
\bibitem [{\citenamefont {Hirschfeld}\ and\ \citenamefont
  {Goldenfeld}(1993{\natexlab{a}})}]{PhysRevB.48.4219}%
  \BibitemOpen
  \bibfield  {author} {\bibinfo {author} {\bibfnamefont {P.~J.}\ \bibnamefont
  {Hirschfeld}}\ and\ \bibinfo {author} {\bibfnamefont {N.}~\bibnamefont
  {Goldenfeld}},\ }\bibfield  {title} {\bibinfo {title} {Effect of strong
  scattering on the low-temperature penetration depth of a d-wave
  superconductor},\ }\href {https://doi.org/10.1103/PhysRevB.48.4219}
  {\bibfield  {journal} {\bibinfo  {journal} {Phys. Rev. B}\ }\textbf {\bibinfo
  {volume} {48}},\ \bibinfo {pages} {4219} (\bibinfo {year}
  {1993}{\natexlab{a}})}\BibitemShut {NoStop}%
\bibitem [{\citenamefont {Lee-Hone}\ \emph {et~al.}(2017)\citenamefont
  {Lee-Hone}, \citenamefont {Dodge},\ and\ \citenamefont
  {Broun}}]{PhysRevB.96.024501}%
  \BibitemOpen
  \bibfield  {author} {\bibinfo {author} {\bibfnamefont {N.~R.}\ \bibnamefont
  {Lee-Hone}}, \bibinfo {author} {\bibfnamefont {J.~S.}\ \bibnamefont
  {Dodge}},\ and\ \bibinfo {author} {\bibfnamefont {D.~M.}\ \bibnamefont
  {Broun}},\ }\bibfield  {title} {\bibinfo {title} {Disorder and superfluid
  density in overdoped cuprate superconductors},\ }\href
  {https://doi.org/10.1103/PhysRevB.96.024501} {\bibfield  {journal} {\bibinfo
  {journal} {Phys. Rev. B}\ }\textbf {\bibinfo {volume} {96}},\ \bibinfo
  {pages} {024501} (\bibinfo {year} {2017})}\BibitemShut {NoStop}%
\bibitem [{\citenamefont {Lee-Hone}\ \emph {et~al.}(2018)\citenamefont
  {Lee-Hone}, \citenamefont {Mishra}, \citenamefont {Broun},\ and\
  \citenamefont {Hirschfeld}}]{PhysRevB.98.054506}%
  \BibitemOpen
  \bibfield  {author} {\bibinfo {author} {\bibfnamefont {N.~R.}\ \bibnamefont
  {Lee-Hone}}, \bibinfo {author} {\bibfnamefont {V.}~\bibnamefont {Mishra}},
  \bibinfo {author} {\bibfnamefont {D.~M.}\ \bibnamefont {Broun}},\ and\
  \bibinfo {author} {\bibfnamefont {P.~J.}\ \bibnamefont {Hirschfeld}},\
  }\bibfield  {title} {\bibinfo {title} {Optical conductivity of overdoped
  cuprate superconductors: Application to {La}$_{2-x}${Sr}$_{x}${CuO}$_{4}$},\
  }\href {https://doi.org/10.1103/PhysRevB.98.054506} {\bibfield  {journal}
  {\bibinfo  {journal} {Phys. Rev. B}\ }\textbf {\bibinfo {volume} {98}},\
  \bibinfo {pages} {054506} (\bibinfo {year} {2018})}\BibitemShut {NoStop}%
\bibitem [{\citenamefont {Lee-Hone}\ \emph {et~al.}(2020)\citenamefont
  {Lee-Hone}, \citenamefont {\"Ozdemir}, \citenamefont {Mishra}, \citenamefont
  {Broun},\ and\ \citenamefont {Hirschfeld}}]{PhysRevResearch.2.013228}%
  \BibitemOpen
  \bibfield  {author} {\bibinfo {author} {\bibfnamefont {N.~R.}\ \bibnamefont
  {Lee-Hone}}, \bibinfo {author} {\bibfnamefont {H.~U.}\ \bibnamefont
  {\"Ozdemir}}, \bibinfo {author} {\bibfnamefont {V.}~\bibnamefont {Mishra}},
  \bibinfo {author} {\bibfnamefont {D.~M.}\ \bibnamefont {Broun}},\ and\
  \bibinfo {author} {\bibfnamefont {P.~J.}\ \bibnamefont {Hirschfeld}},\
  }\bibfield  {title} {\bibinfo {title} {Low energy phenomenology of the
  overdoped cuprates: Viability of the {Landau-BCS} paradigm},\ }\href
  {https://doi.org/10.1103/PhysRevResearch.2.013228} {\bibfield  {journal}
  {\bibinfo  {journal} {Phys. Rev. Research}\ }\textbf {\bibinfo {volume}
  {2}},\ \bibinfo {pages} {013228} (\bibinfo {year} {2020})}\BibitemShut
  {NoStop}%
\bibitem [{\citenamefont {Özdemir}\ \emph {et~al.}(2022)\citenamefont
  {Özdemir}, \citenamefont {Mishra}, \citenamefont {Lee-Hone}, \citenamefont
  {Kong}, \citenamefont {Berlijn}, \citenamefont {Broun},\ and\ \citenamefont
  {Hirschfeld}}]{https://doi.org/10.48550/arxiv.2206.01348}%
  \BibitemOpen
  \bibfield  {author} {\bibinfo {author} {\bibfnamefont {H.~U.}\ \bibnamefont
  {Özdemir}}, \bibinfo {author} {\bibfnamefont {V.}~\bibnamefont {Mishra}},
  \bibinfo {author} {\bibfnamefont {N.~R.}\ \bibnamefont {Lee-Hone}}, \bibinfo
  {author} {\bibfnamefont {X.}~\bibnamefont {Kong}}, \bibinfo {author}
  {\bibfnamefont {T.}~\bibnamefont {Berlijn}}, \bibinfo {author} {\bibfnamefont
  {D.~M.}\ \bibnamefont {Broun}},\ and\ \bibinfo {author} {\bibfnamefont
  {P.~J.}\ \bibnamefont {Hirschfeld}},\ }\href
  {https://doi.org/10.48550/ARXIV.2206.01348} {\bibinfo {title} {From {Mott} to
  not: effect of out-of-plane dopants on superfluid density of overdoped
  cuprates}} (\bibinfo {year} {2022})\BibitemShut {NoStop}%
\bibitem [{\citenamefont {{Li}}\ \emph {et~al.}(2022)\citenamefont {{Li}},
  \citenamefont {{Sapkota}}, \citenamefont {{Lozano}}, \citenamefont {{Du}},
  \citenamefont {{Li}}, \citenamefont {{Wu}}, \citenamefont {{Kundu}},
  \citenamefont {{Winn}}, \citenamefont {{Chi}}, \citenamefont {{Matsuda}},
  \citenamefont {{Frontzek}}, \citenamefont {{Bozovic}}, \citenamefont
  {{Pasupathy}}, \citenamefont {{Drozdov}}, \citenamefont {{Fujita}},
  \citenamefont {{Gu}}, \citenamefont {{Zaliznyak}}, \citenamefont {{Li}},\
  and\ \citenamefont {{Tranquada}}}]{Tranquada2022}%
  \BibitemOpen
  \bibfield  {author} {\bibinfo {author} {\bibfnamefont {Y.}~\bibnamefont
  {{Li}}}, \bibinfo {author} {\bibfnamefont {A.}~\bibnamefont {{Sapkota}}},
  \bibinfo {author} {\bibfnamefont {P.~M.}\ \bibnamefont {{Lozano}}}, \bibinfo
  {author} {\bibfnamefont {Z.}~\bibnamefont {{Du}}}, \bibinfo {author}
  {\bibfnamefont {H.}~\bibnamefont {{Li}}}, \bibinfo {author} {\bibfnamefont
  {Z.}~\bibnamefont {{Wu}}}, \bibinfo {author} {\bibfnamefont {A.~K.}\
  \bibnamefont {{Kundu}}}, \bibinfo {author} {\bibfnamefont {B.~L.}\
  \bibnamefont {{Winn}}}, \bibinfo {author} {\bibfnamefont {S.}~\bibnamefont
  {{Chi}}}, \bibinfo {author} {\bibfnamefont {M.}~\bibnamefont {{Matsuda}}},
  \bibinfo {author} {\bibfnamefont {M.}~\bibnamefont {{Frontzek}}}, \bibinfo
  {author} {\bibfnamefont {I.}~\bibnamefont {{Bozovic}}}, \bibinfo {author}
  {\bibfnamefont {A.~N.}\ \bibnamefont {{Pasupathy}}}, \bibinfo {author}
  {\bibfnamefont {I.~K.}\ \bibnamefont {{Drozdov}}}, \bibinfo {author}
  {\bibfnamefont {K.}~\bibnamefont {{Fujita}}}, \bibinfo {author}
  {\bibfnamefont {G.~D.}\ \bibnamefont {{Gu}}}, \bibinfo {author}
  {\bibfnamefont {I.}~\bibnamefont {{Zaliznyak}}}, \bibinfo {author}
  {\bibfnamefont {Q.}~\bibnamefont {{Li}}},\ and\ \bibinfo {author}
  {\bibfnamefont {J.~M.}\ \bibnamefont {{Tranquada}}},\ }\bibfield  {title}
  {\bibinfo {title} {{Strongly-overdoped La$_{2-x}$Sr$_x$CuO$_4$: Evidence for
  Josephson-coupled grains of strongly-correlated superconductor}},\
  }\href@noop {} {\bibfield  {journal} {\bibinfo  {journal} {arXiv e-prints}\
  ,\ \bibinfo {eid} {arXiv:2205.01702}} (\bibinfo {year} {2022})},\ \Eprint
  {https://arxiv.org/abs/2205.01702} {arXiv:2205.01702 [cond-mat.supr-con]}
  \BibitemShut {NoStop}%
\bibitem [{\citenamefont {{Tromp}}\ \emph {et~al.}(2022)\citenamefont
  {{Tromp}}, \citenamefont {{Benschop}}, \citenamefont {{Ge}}, \citenamefont
  {{Battisti}}, \citenamefont {{Bastiaans}}, \citenamefont {{Chatzopoulos}},
  \citenamefont {{Vervloet}}, \citenamefont {{Smit}}, \citenamefont {{van
  Heumen}}, \citenamefont {{Golden}}, \citenamefont {{Huang}}, \citenamefont
  {{Kondo}}, \citenamefont {{Yin}}, \citenamefont {{Hoffman}}, \citenamefont
  {{Sulangi}}, \citenamefont {{Zaanen}},\ and\ \citenamefont
  {{Allan}}}]{MilanAllan2022}%
  \BibitemOpen
  \bibfield  {author} {\bibinfo {author} {\bibfnamefont {W.~O.}\ \bibnamefont
  {{Tromp}}}, \bibinfo {author} {\bibfnamefont {T.}~\bibnamefont {{Benschop}}},
  \bibinfo {author} {\bibfnamefont {J.-F.}\ \bibnamefont {{Ge}}}, \bibinfo
  {author} {\bibfnamefont {I.}~\bibnamefont {{Battisti}}}, \bibinfo {author}
  {\bibfnamefont {K.~M.}\ \bibnamefont {{Bastiaans}}}, \bibinfo {author}
  {\bibfnamefont {D.}~\bibnamefont {{Chatzopoulos}}}, \bibinfo {author}
  {\bibfnamefont {A.}~\bibnamefont {{Vervloet}}}, \bibinfo {author}
  {\bibfnamefont {S.}~\bibnamefont {{Smit}}}, \bibinfo {author} {\bibfnamefont
  {E.}~\bibnamefont {{van Heumen}}}, \bibinfo {author} {\bibfnamefont {M.~S.}\
  \bibnamefont {{Golden}}}, \bibinfo {author} {\bibfnamefont {Y.}~\bibnamefont
  {{Huang}}}, \bibinfo {author} {\bibfnamefont {T.}~\bibnamefont {{Kondo}}},
  \bibinfo {author} {\bibfnamefont {Y.}~\bibnamefont {{Yin}}}, \bibinfo
  {author} {\bibfnamefont {J.~E.}\ \bibnamefont {{Hoffman}}}, \bibinfo {author}
  {\bibfnamefont {M.~A.}\ \bibnamefont {{Sulangi}}}, \bibinfo {author}
  {\bibfnamefont {J.}~\bibnamefont {{Zaanen}}},\ and\ \bibinfo {author}
  {\bibfnamefont {M.~P.}\ \bibnamefont {{Allan}}},\ }\bibfield  {title}
  {\bibinfo {title} {{Puddle formation, persistent gaps, and non-mean-field
  breakdown of superconductivity in overdoped (Pb,Bi)2Sr2CuO6+$\delta$}},\
  }\href@noop {} {\bibfield  {journal} {\bibinfo  {journal} {arXiv e-prints}\
  ,\ \bibinfo {eid} {arXiv:2205.09740}} (\bibinfo {year} {2022})},\ \Eprint
  {https://arxiv.org/abs/2205.09740} {arXiv:2205.09740 [cond-mat.supr-con]}
  \BibitemShut {NoStop}%
\bibitem [{\citenamefont {Benfatto}\ \emph {et~al.}(2001)\citenamefont
  {Benfatto}, \citenamefont {Toschi}, \citenamefont {Caprara},\ and\
  \citenamefont {Castellani}}]{Benfatto2001}%
  \BibitemOpen
  \bibfield  {author} {\bibinfo {author} {\bibfnamefont {L.}~\bibnamefont
  {Benfatto}}, \bibinfo {author} {\bibfnamefont {A.}~\bibnamefont {Toschi}},
  \bibinfo {author} {\bibfnamefont {S.}~\bibnamefont {Caprara}},\ and\ \bibinfo
  {author} {\bibfnamefont {C.}~\bibnamefont {Castellani}},\ }\bibfield  {title}
  {\bibinfo {title} {Phase fluctuations in superconductors: From galilean
  invariant to quantum $\mathrm{XY}$ models},\ }\href
  {https://doi.org/10.1103/PhysRevB.64.140506} {\bibfield  {journal} {\bibinfo
  {journal} {Phys. Rev. B}\ }\textbf {\bibinfo {volume} {64}},\ \bibinfo
  {pages} {140506} (\bibinfo {year} {2001})}\BibitemShut {NoStop}%
\bibitem [{\citenamefont {Brei\o{}}\ \emph {et~al.}(2022)\citenamefont
  {Brei\o{}}, \citenamefont {Hirschfeld},\ and\ \citenamefont
  {Andersen}}]{Breio2022}%
  \BibitemOpen
  \bibfield  {author} {\bibinfo {author} {\bibfnamefont {C.~N.}\ \bibnamefont
  {Brei\o{}}}, \bibinfo {author} {\bibfnamefont {P.~J.}\ \bibnamefont
  {Hirschfeld}},\ and\ \bibinfo {author} {\bibfnamefont {B.~M.}\ \bibnamefont
  {Andersen}},\ }\bibfield  {title} {\bibinfo {title} {Supercurrents and
  spontaneous time-reversal symmetry breaking by nonmagnetic disorder in
  unconventional superconductors},\ }\href
  {https://doi.org/10.1103/PhysRevB.105.014504} {\bibfield  {journal} {\bibinfo
   {journal} {Phys. Rev. B}\ }\textbf {\bibinfo {volume} {105}},\ \bibinfo
  {pages} {014504} (\bibinfo {year} {2022})}\BibitemShut {NoStop}%
\bibitem [{\citenamefont {Nunner}\ \emph {et~al.}(2005)\citenamefont {Nunner},
  \citenamefont {Andersen}, \citenamefont {Melikyan},\ and\ \citenamefont
  {Hirschfeld}}]{Nunner2005}%
  \BibitemOpen
  \bibfield  {author} {\bibinfo {author} {\bibfnamefont {T.~S.}\ \bibnamefont
  {Nunner}}, \bibinfo {author} {\bibfnamefont {B.~M.}\ \bibnamefont
  {Andersen}}, \bibinfo {author} {\bibfnamefont {A.}~\bibnamefont {Melikyan}},\
  and\ \bibinfo {author} {\bibfnamefont {P.~J.}\ \bibnamefont {Hirschfeld}},\
  }\bibfield  {title} {\bibinfo {title} {Dopant-modulated pair interaction in
  cuprate superconductors},\ }\href
  {https://doi.org/10.1103/PhysRevLett.95.177003} {\bibfield  {journal}
  {\bibinfo  {journal} {Phys. Rev. Lett.}\ }\textbf {\bibinfo {volume} {95}},\
  \bibinfo {pages} {177003} (\bibinfo {year} {2005})}\BibitemShut {NoStop}%
\bibitem [{\citenamefont {Atkinson}\ \emph
  {et~al.}(2000{\natexlab{a}})\citenamefont {Atkinson}, \citenamefont
  {Hirschfeld},\ and\ \citenamefont {MacDonald}}]{Atkinson2000}%
  \BibitemOpen
  \bibfield  {author} {\bibinfo {author} {\bibfnamefont {W.~A.}\ \bibnamefont
  {Atkinson}}, \bibinfo {author} {\bibfnamefont {P.~J.}\ \bibnamefont
  {Hirschfeld}},\ and\ \bibinfo {author} {\bibfnamefont {A.~H.}\ \bibnamefont
  {MacDonald}},\ }\bibfield  {title} {\bibinfo {title} {Gap inhomogeneities and
  the density of states in disordered d-wave superconductors},\ }\href
  {https://doi.org/10.1103/PhysRevLett.85.3922} {\bibfield  {journal} {\bibinfo
   {journal} {Phys. Rev. Lett.}\ }\textbf {\bibinfo {volume} {85}},\ \bibinfo
  {pages} {3922} (\bibinfo {year} {2000}{\natexlab{a}})}\BibitemShut {NoStop}%
\bibitem [{\citenamefont {Ghosal}\ \emph {et~al.}(2000)\citenamefont {Ghosal},
  \citenamefont {Randeria},\ and\ \citenamefont {Trivedi}}]{Ghosal2000}%
  \BibitemOpen
  \bibfield  {author} {\bibinfo {author} {\bibfnamefont {A.}~\bibnamefont
  {Ghosal}}, \bibinfo {author} {\bibfnamefont {M.}~\bibnamefont {Randeria}},\
  and\ \bibinfo {author} {\bibfnamefont {N.}~\bibnamefont {Trivedi}},\
  }\bibfield  {title} {\bibinfo {title} {Spatial inhomogeneities in disordered
  d-wave superconductors},\ }\href {https://doi.org/10.1103/PhysRevB.63.020505}
  {\bibfield  {journal} {\bibinfo  {journal} {Phys. Rev. B}\ }\textbf {\bibinfo
  {volume} {63}},\ \bibinfo {pages} {020505} (\bibinfo {year}
  {2000})}\BibitemShut {NoStop}%
\bibitem [{\citenamefont {Atkinson}\ \emph
  {et~al.}(2000{\natexlab{b}})\citenamefont {Atkinson}, \citenamefont
  {Hirschfeld}, \citenamefont {MacDonald},\ and\ \citenamefont
  {Ziegler}}]{Atkinson2000B}%
  \BibitemOpen
  \bibfield  {author} {\bibinfo {author} {\bibfnamefont {W.~A.}\ \bibnamefont
  {Atkinson}}, \bibinfo {author} {\bibfnamefont {P.~J.}\ \bibnamefont
  {Hirschfeld}}, \bibinfo {author} {\bibfnamefont {A.~H.}\ \bibnamefont
  {MacDonald}},\ and\ \bibinfo {author} {\bibfnamefont {K.}~\bibnamefont
  {Ziegler}},\ }\bibfield  {title} {\bibinfo {title} {Details of disorder
  matter in 2d $\mathit{d}$-wave superconductors},\ }\href
  {https://doi.org/10.1103/PhysRevLett.85.3926} {\bibfield  {journal} {\bibinfo
   {journal} {Phys. Rev. Lett.}\ }\textbf {\bibinfo {volume} {85}},\ \bibinfo
  {pages} {3926} (\bibinfo {year} {2000}{\natexlab{b}})}\BibitemShut {NoStop}%
\bibitem [{\citenamefont {Yashenkin}\ \emph {et~al.}(2001)\citenamefont
  {Yashenkin}, \citenamefont {Atkinson}, \citenamefont {Gornyi}, \citenamefont
  {Hirschfeld},\ and\ \citenamefont {Khveshchenko}}]{Yashenkin2001}%
  \BibitemOpen
  \bibfield  {author} {\bibinfo {author} {\bibfnamefont {A.~G.}\ \bibnamefont
  {Yashenkin}}, \bibinfo {author} {\bibfnamefont {W.~A.}\ \bibnamefont
  {Atkinson}}, \bibinfo {author} {\bibfnamefont {I.~V.}\ \bibnamefont
  {Gornyi}}, \bibinfo {author} {\bibfnamefont {P.~J.}\ \bibnamefont
  {Hirschfeld}},\ and\ \bibinfo {author} {\bibfnamefont {D.~V.}\ \bibnamefont
  {Khveshchenko}},\ }\bibfield  {title} {\bibinfo {title} {Nesting symmetries
  and diffusion in disordered $\mathit{d}$-wave superconductors},\ }\href
  {https://doi.org/10.1103/PhysRevLett.86.5982} {\bibfield  {journal} {\bibinfo
   {journal} {Phys. Rev. Lett.}\ }\textbf {\bibinfo {volume} {86}},\ \bibinfo
  {pages} {5982} (\bibinfo {year} {2001})}\BibitemShut {NoStop}%
\bibitem [{\citenamefont {Atkinson}\ \emph {et~al.}(2003)\citenamefont
  {Atkinson}, \citenamefont {Hirschfeld},\ and\ \citenamefont
  {Zhu}}]{Atkinson2003}%
  \BibitemOpen
  \bibfield  {author} {\bibinfo {author} {\bibfnamefont {W.~A.}\ \bibnamefont
  {Atkinson}}, \bibinfo {author} {\bibfnamefont {P.~J.}\ \bibnamefont
  {Hirschfeld}},\ and\ \bibinfo {author} {\bibfnamefont {L.}~\bibnamefont
  {Zhu}},\ }\bibfield  {title} {\bibinfo {title} {Quantum interference in
  nested d-wave superconductors: A real-space perspective},\ }\href
  {https://doi.org/10.1103/PhysRevB.68.054501} {\bibfield  {journal} {\bibinfo
  {journal} {Phys. Rev. B}\ }\textbf {\bibinfo {volume} {68}},\ \bibinfo
  {pages} {054501} (\bibinfo {year} {2003})}\BibitemShut {NoStop}%
\bibitem [{\citenamefont {Garg}\ \emph {et~al.}(2008)\citenamefont {Garg},
  \citenamefont {Randeria},\ and\ \citenamefont {Trivedi}}]{Garg2008}%
  \BibitemOpen
  \bibfield  {author} {\bibinfo {author} {\bibfnamefont {A.}~\bibnamefont
  {Garg}}, \bibinfo {author} {\bibfnamefont {M.}~\bibnamefont {Randeria}},\
  and\ \bibinfo {author} {\bibfnamefont {N.}~\bibnamefont {Trivedi}},\
  }\bibfield  {title} {\bibinfo {title} {Strong correlations make
  high-temperature superconductors robust against disorder},\ }\href
  {https://doi.org/10.1038/nphys1026} {\bibfield  {journal} {\bibinfo
  {journal} {Nature Physics}\ }\textbf {\bibinfo {volume} {4}},\ \bibinfo
  {pages} {762} (\bibinfo {year} {2008})}\BibitemShut {NoStop}%
\bibitem [{\citenamefont {Hirschfeld}\ and\ \citenamefont
  {Atkinson}(2002)}]{PJH_WAA_Jltp}%
  \BibitemOpen
  \bibfield  {author} {\bibinfo {author} {\bibfnamefont {P.~J.}\ \bibnamefont
  {Hirschfeld}}\ and\ \bibinfo {author} {\bibfnamefont {W.~A.}\ \bibnamefont
  {Atkinson}},\ }\bibfield  {title} {\bibinfo {title} {$\pi$ {\`a} la node:
  Disordered d-wave superconductors in two dimensions for the random masses},\
  }\href {https://doi.org/10.1023/A:1013838523587} {\bibfield  {journal}
  {\bibinfo  {journal} {Journal of Low Temperature Physics}\ }\textbf {\bibinfo
  {volume} {126}},\ \bibinfo {pages} {881} (\bibinfo {year}
  {2002})}\BibitemShut {NoStop}%
\bibitem [{\citenamefont {Atkinson}(2007)}]{PhysRevB.75.024510}%
  \BibitemOpen
  \bibfield  {author} {\bibinfo {author} {\bibfnamefont {W.~A.}\ \bibnamefont
  {Atkinson}},\ }\bibfield  {title} {\bibinfo {title} {Superfluid suppression
  in $d$-wave superconductors due to disordered magnetism},\ }\href
  {https://doi.org/10.1103/PhysRevB.75.024510} {\bibfield  {journal} {\bibinfo
  {journal} {Phys. Rev. B}\ }\textbf {\bibinfo {volume} {75}},\ \bibinfo
  {pages} {024510} (\bibinfo {year} {2007})}\BibitemShut {NoStop}%
\bibitem [{\citenamefont {Scalapino}\ \emph {et~al.}(1993)\citenamefont
  {Scalapino}, \citenamefont {White},\ and\ \citenamefont
  {Zhang}}]{PhysRevB.47.7995}%
  \BibitemOpen
  \bibfield  {author} {\bibinfo {author} {\bibfnamefont {D.~J.}\ \bibnamefont
  {Scalapino}}, \bibinfo {author} {\bibfnamefont {S.~R.}\ \bibnamefont
  {White}},\ and\ \bibinfo {author} {\bibfnamefont {S.}~\bibnamefont {Zhang}},\
  }\bibfield  {title} {\bibinfo {title} {Insulator, metal, or superconductor:
  The criteria},\ }\href {https://doi.org/10.1103/PhysRevB.47.7995} {\bibfield
  {journal} {\bibinfo  {journal} {Phys. Rev. B}\ }\textbf {\bibinfo {volume}
  {47}},\ \bibinfo {pages} {7995} (\bibinfo {year} {1993})}\BibitemShut
  {NoStop}%
\bibitem [{\citenamefont {Skalski}\ \emph {et~al.}(1964)\citenamefont
  {Skalski}, \citenamefont {Betbeder-Matibet},\ and\ \citenamefont
  {Weiss}}]{Skalski1964}%
  \BibitemOpen
  \bibfield  {author} {\bibinfo {author} {\bibfnamefont {S.}~\bibnamefont
  {Skalski}}, \bibinfo {author} {\bibfnamefont {O.}~\bibnamefont
  {Betbeder-Matibet}},\ and\ \bibinfo {author} {\bibfnamefont {P.~R.}\
  \bibnamefont {Weiss}},\ }\bibfield  {title} {\bibinfo {title} {Properties of
  superconducting alloys containing paramagnetic impurities},\ }\href
  {https://doi.org/10.1103/PhysRev.136.A1500} {\bibfield  {journal} {\bibinfo
  {journal} {Phys. Rev.}\ }\textbf {\bibinfo {volume} {136}},\ \bibinfo {pages}
  {A1500} (\bibinfo {year} {1964})}\BibitemShut {NoStop}%
\bibitem [{\citenamefont {Dodaro}\ and\ \citenamefont
  {Kivelson}(2018)}]{Dodaro2018}%
  \BibitemOpen
  \bibfield  {author} {\bibinfo {author} {\bibfnamefont {J.~F.}\ \bibnamefont
  {Dodaro}}\ and\ \bibinfo {author} {\bibfnamefont {S.~A.}\ \bibnamefont
  {Kivelson}},\ }\bibfield  {title} {\bibinfo {title} {Generalization of
  anderson's theorem for disordered superconductors},\ }\href
  {https://doi.org/10.1103/PhysRevB.98.174503} {\bibfield  {journal} {\bibinfo
  {journal} {Phys. Rev. B}\ }\textbf {\bibinfo {volume} {98}},\ \bibinfo
  {pages} {174503} (\bibinfo {year} {2018})}\BibitemShut {NoStop}%
\bibitem [{\citenamefont {Hardy}\ \emph {et~al.}(1993)\citenamefont {Hardy},
  \citenamefont {Bonn}, \citenamefont {Morgan}, \citenamefont {Liang},\ and\
  \citenamefont {Zhang}}]{Hardy1993}%
  \BibitemOpen
  \bibfield  {author} {\bibinfo {author} {\bibfnamefont {W.~N.}\ \bibnamefont
  {Hardy}}, \bibinfo {author} {\bibfnamefont {D.~A.}\ \bibnamefont {Bonn}},
  \bibinfo {author} {\bibfnamefont {D.~C.}\ \bibnamefont {Morgan}}, \bibinfo
  {author} {\bibfnamefont {R.}~\bibnamefont {Liang}},\ and\ \bibinfo {author}
  {\bibfnamefont {K.}~\bibnamefont {Zhang}},\ }\bibfield  {title} {\bibinfo
  {title} {Precision measurements of the temperature dependence of
  \ensuremath{\lambda} in
  {${\mathrm{YBa}}_{2}$${\mathrm{Cu}}_{3}$${\mathrm{O}}_{6.95}$}: Strong
  evidence for nodes in the gap function},\ }\href
  {https://doi.org/10.1103/PhysRevLett.70.3999} {\bibfield  {journal} {\bibinfo
   {journal} {Phys. Rev. Lett.}\ }\textbf {\bibinfo {volume} {70}},\ \bibinfo
  {pages} {3999} (\bibinfo {year} {1993})}\BibitemShut {NoStop}%
\bibitem [{\citenamefont {Achkir}\ \emph {et~al.}(1993)\citenamefont {Achkir},
  \citenamefont {Poirier}, \citenamefont {Bonn}, \citenamefont {Liang},\ and\
  \citenamefont {Hardy}}]{BonnHardy1993Zn}%
  \BibitemOpen
  \bibfield  {author} {\bibinfo {author} {\bibfnamefont {D.}~\bibnamefont
  {Achkir}}, \bibinfo {author} {\bibfnamefont {M.}~\bibnamefont {Poirier}},
  \bibinfo {author} {\bibfnamefont {D.~A.}\ \bibnamefont {Bonn}}, \bibinfo
  {author} {\bibfnamefont {R.}~\bibnamefont {Liang}},\ and\ \bibinfo {author}
  {\bibfnamefont {W.~N.}\ \bibnamefont {Hardy}},\ }\bibfield  {title} {\bibinfo
  {title} {Temperature dependence of the in-plane penetration depth of
  {${\mathrm{YBa}}_{2}{\mathrm{Cu}}_{3}{\mathrm{O}}_{6.95}$} and
  {${\mathrm{YBa}}_{2}$(${\mathrm{Cu}}_{0.9985}$${\mathrm{Zn}}_{0.0015}$${)}_{3}$${\mathrm{O}}_{6.95}$}
  crystals from t to ${\mathit{t}}^{2}$},\ }\href
  {https://doi.org/10.1103/PhysRevB.48.13184} {\bibfield  {journal} {\bibinfo
  {journal} {Phys. Rev. B}\ }\textbf {\bibinfo {volume} {48}},\ \bibinfo
  {pages} {13184} (\bibinfo {year} {1993})}\BibitemShut {NoStop}%
\bibitem [{\citenamefont {Prohammer}\ and\ \citenamefont
  {Carbotte}(1991)}]{prohammer1991}%
  \BibitemOpen
  \bibfield  {author} {\bibinfo {author} {\bibfnamefont {M.}~\bibnamefont
  {Prohammer}}\ and\ \bibinfo {author} {\bibfnamefont {J.~P.}\ \bibnamefont
  {Carbotte}},\ }\bibfield  {title} {\bibinfo {title} {{London penetration
  depth of $d$-wave\ superconductors}},\ }\href@noop {} {\bibfield  {journal}
  {\bibinfo  {journal} {Phys.\ Rev.\ B}\ }\textbf {\bibinfo {volume} {43}},\
  \bibinfo {pages} {5370} (\bibinfo {year} {1991})}\BibitemShut {NoStop}%
\bibitem [{\citenamefont {Hirschfeld}\ and\ \citenamefont
  {Goldenfeld}(1993{\natexlab{b}})}]{Hirschfeld:1993cka}%
  \BibitemOpen
  \bibfield  {author} {\bibinfo {author} {\bibfnamefont {P.~J.}\ \bibnamefont
  {Hirschfeld}}\ and\ \bibinfo {author} {\bibfnamefont {N.}~\bibnamefont
  {Goldenfeld}},\ }\bibfield  {title} {\bibinfo {title} {Effect of strong
  scattering on the low-temperature penetration depth of a d-wave
  superconductor},\ }\href {https://doi.org/10.1103/PhysRevB.48.4219}
  {\bibfield  {journal} {\bibinfo  {journal} {Phys. Rev. B}\ }\textbf {\bibinfo
  {volume} {48}},\ \bibinfo {pages} {4219} (\bibinfo {year}
  {1993}{\natexlab{b}})}\BibitemShut {NoStop}%
\bibitem [{\citenamefont {Sheehy}\ \emph {et~al.}(2004)\citenamefont {Sheehy},
  \citenamefont {Davis},\ and\ \citenamefont {Franz}}]{PhysRevB.70.054510}%
  \BibitemOpen
  \bibfield  {author} {\bibinfo {author} {\bibfnamefont {D.~E.}\ \bibnamefont
  {Sheehy}}, \bibinfo {author} {\bibfnamefont {T.~P.}\ \bibnamefont {Davis}},\
  and\ \bibinfo {author} {\bibfnamefont {M.}~\bibnamefont {Franz}},\ }\bibfield
   {title} {\bibinfo {title} {Unified theory of the $\mathrm{ab}$-plane and
  c-axis penetration depths of underdoped cuprates},\ }\href
  {https://doi.org/10.1103/PhysRevB.70.054510} {\bibfield  {journal} {\bibinfo
  {journal} {Phys. Rev. B}\ }\textbf {\bibinfo {volume} {70}},\ \bibinfo
  {pages} {054510} (\bibinfo {year} {2004})}\BibitemShut {NoStop}%
\end{thebibliography}%
%\bibliography{overdoped.bib}
\end{document}